\documentclass[structabstract]{aa}
\usepackage{graphics}
\usepackage{graphicx}
\usepackage{txfonts}
\usepackage{multirow}  
\usepackage{enumerate}
\usepackage{hyperref} 
\usepackage{color}
\definecolor{darkblue}{rgb}{0,0,0.1}
\definecolor{red}{rgb}{1,0,0}
\hypersetup{pdftex=true, colorlinks=true, breaklinks=true,
linkcolor=darkblue, menucolor=darkblue, pagecolor=darkblue,
urlcolor=darkblue}
\newcommand{\V}[1]{{\textbf{{\em#1}}}}
\newcommand{\TB}[1]{{{{#1}}}}

\begin{document}
   \title{Optimization of cw sodium laser guide star efficiency}


   \author{R. Holzl\"ohner\inst{1}
      \and
      S. M. Rochester\inst{2}
	\and
	D. Bonaccini Calia\inst{1}
	\and
	D. Budker\inst{2}
	\and
	J. M. Higbie\inst{3}
	\and
	W. Hackenberg\inst{1}
          }

   \institute{Laser Systems Department, European Southern Observatory
  (ESO), Karl-Schwarzschild-Str. 2, D-85748 Garching b.~M\"unchen, Germany,
  \href{http://www.eso.org/sci/facilities/develop/lgsf/}{http://www.eso.org/sci/facilities/develop/lgsf/}\\
	\email{rholzloe[at]eso.org, dbonacci[at]eso.org, whackenb[at]eso.org}
         \and
  University of California Berkeley, Department of Physics,
  Berkeley, CA 94720-7300, USA\\ \email{simonr[at]berkeley.edu, budker[at].berkeley.edu}
	\and
  Bucknell University, Department of Physics, 701 Moore Avenue,
  Lewisburg, PA 17837, USA \\ \email{james.higbie[at]bucknell.edu}}

   \date{Received xxx; accepted xxx}


  \abstract
   {Sodium laser guide stars (LGS) are about to enter a new range of
 laser powers. Previous theoretical and numerical methods
are inadequate for accurate computations of the return flux and hence for
 the design of the next-generation LGS systems.}
   {We {\TB{numerically}} optimize the cw (continuous wave) laser format, in particular the
light polarization and spectrum.}
   {Using Bloch equations, we simulate the mesospheric sodium atoms,
   including Doppler broadening, saturation, collisional relaxation,
   Larmor precession, and recoil, taking into account all 24 sodium
   hyperfine states and on the order of 100 velocity groups.}
   {LGS return flux is limited by `three evils': Larmor precession due to
the geomagnetic field, atomic recoil due to radiation pressure,
and transition saturation. We study their
impacts and show that the return flux can be boosted by repumping
(simultaneous excitation of the sodium D$_2$a and
D$_2$b lines with 10--20\,\% of the laser power in the latter).
}
  {We strongly recommend the use of circularly polarized lasers and
repumping. As a rule of thumb, the bandwidth of laser radiation in MHz (at
each line) should approximately equal the launched
laser power in Watts divided by six, assuming a diffraction-limited spot size.}

   \keywords{sodium laser guide stars --
                resonance fluorescence --
		Bloch equations --
                atomic physics}

   \maketitle
%

\section{Introduction}

Laser guide stars (LGS) are becoming essential in providing artificial beacons for adaptive optics (AO) in large telescopes (Hubin \cite{Hubin09}). The generation of 8--10\,m class telescopes such as the Very Large Telescope (VLT) on Cerro Paranal, Chile, or the Keck telescopes on Mauna Kea, has been retrofitted with LGS and is still operated in many observing programs with natural guide stars or no AO at all. The upcoming 30+\,m telescopes such as the 42-m European Extremely Large Telescope (E-ELT) or the Thirty Meter Telescope (TMT), by contrast, are being designed from the start as adaptive telescopes and will require LGS in nearly all of their science operation. At the same time, sensing subaperture sizes of 20--50\,cm and near-kHz AO frame rates, designed to provide high Strehl ratios in the infrared or even in the visible, require unprecedented LGS brightness and hence laser power.

Sodium (Na) LGS at 589\,nm are most commonly used because of the large fluorescence \TB{cross section-abundance product of ca.~$10^{-11}$\,cm$^2 \times 4\!\times\!10^{13}$\,m$^{-2}$, a fluorescence wavelength in the visible,} and the high altitude of the sodium layer around 80--100\,km compared e.g. to Rayleigh LGS, (Happer \cite{Happer94}, Table~1), which allows one to sense a large fraction of the turbulent atmosphere column above the telescope with a small number of guide stars. Unfortunately, powerful diffraction-limited laser beams at 589\,nm are quite expensive to produce due to the lack of solid-state materials that amplify 589\,nm or 1178\,nm light. Recently, ESO has demonstrated a frequency-doubled narrow-band Raman fiber laser emitting 25\,W at 589\,nm (Taylor \cite{TaylorCLEO09}), and we expect this technology to significantly improve the experimental situation.

Careful dimensioning of the required laser power and optimization of the laser format (spectrum, polarization, spot size) are needed. Numerical simulations are necessary since the experimental situation so far is unsatisfactory: The sodium layer and atmospheric parameters often fluctuate rapidly \TB{(Thomas \cite{Thomas}, Pfrommer \cite{Pfrommer})}, few reliable powerful lasers at 589\,nm have been available up to now, and LGS sky experiments lack commonly agreed measurement standards. Sodium cells cannot easily simulate mesospheric conditions. This work provides optimization rules for the case of continuous wave (cw) lasers over a wide range of laser powers.

Sodium LGS take advantage of the $3^2\!S_{1/2}$--$3^2\!P_{3/2}$ dipole transition, known as the D$_2$ line. The sodium $^2\!S$ ground state consists of two hyperfine multiplets with 8 magnetic substates combined, separated by $1.772$\,GHz, splitting the D$_2$ line into the D$_2$a and D$_2$b transition groups, corresponding respectively to the $F\!=\!1$ and $F\!=\!2$ Na ground states, where $F$ is the total atomic angular momentum quantum number. The four $^2\!P$ multiplets ($F\!=\!0\ldots3$) are separated by only 16, 34, and 60\,MHz, respectively (16 magnetic substates, see Ungar \cite{Ungar}). At mesospheric temperatures near 190\,K, the D$_2$a and D$_2$b lines are Doppler broadened to about 1\,GHz each, giving rise to the characteristic double-hump absorption profile (see for instance Bradley (\cite{Bradley}), Fig.\,11, or Fig.\,\ref{sigma_vs_detuning} of this work). When the D$_2$a transition is excited by circularly polarized light at high irradiance (i.e. optical power density, unit W/m$^2$), a large fraction of atoms is pumped into the $(S,F\!=\!2,m\!=\!\pm2)$ substate, and the atoms cycle on the transition to the $(P,F'\!=\!3,m'\!=\!\pm3)$ substate, with $m$ the magnetic quantum number, so that sodium effectively becomes a two-level system. This situation is twofold desirable because of the large transition cross section and the directional return light peak towards the laser emitter and the telescope. Numerous studies of optical pumping have been conducted, e.g.\ by Happer (\cite{Happer72}) and McClelland (\cite{McClelland}), and the
process is well understood.

The effective sodium return flux depends on the environmental conditions such as collisions with constituent gases, temperature, and the geomagnetic field. At higher irradiance, the atoms are driven away from thermal equilibrium, and it is necessary to take into account effects such as saturation, optical pumping, and recoil caused by radiation pressure. Understanding the complicated interplay of these effects and obtaining quantitative values of the fluorescence efficiency requires numerical simulations. A commonly used method is the solution of the density matrix evolution (Bloch equation) of a multilevel atom.

Milonni and Thode (\cite{Milonni92}) simplified the
D$_2$ scheme to a 2-level Bloch model which they solve in time domain.
Bradley (\cite{Bradley}) simulated the full 24-state density matrix, exciting the sodium by a train of short (nanosecond range) laser pulses like Milonni, using Runge-Kutta integration for one pulse period and exploiting the periodicity. Linear and circular light
polarizations were treated. However, both works neglect the geomagnetic field.
Morris (\cite{Morris}) studied frequency-modulated pulses over a wide
range of pulse durations and linewidths up to 3\,GHz, hence spanning the
entire Doppler broadened D$_2$ line, employing time domain integration, for both linearly and circularly polarized light. In the end of his paper, he briefly estimates the impact of the geomagnetic field, but he does not include Larmor precession terms into his Bloch equations.

To our knowledge, Milonni (\cite{Milonni98,Milonni99}) has published the most advanced and detailed Bloch-equation simulation of sodium LGS to date, later generalized by Telle (\cite{Telle06,Telle08}). In his \cite{Milonni98} article, Milonni treats the cases of laser pulses that are short, comparable, and long compared to the $^2\!P$ lifetime of $\tau\!=\!16.24$\,ns, using numerical solution methods similar to Bradley's. His \cite{Milonni99} publication deals with cw excitation only and introduces spin relaxation and Larmor precession into the Bloch equations for the first time. He shows that the Larmor terms due to the geomagnetic field tend to redistribute the magnetic sublevel populations and hence impede optical pumping. All of the works cited in this and the previous paragraph solve the Bloch equations separately for a number of different sodium velocity classes (100--400) and then perform a weighted average over the results, and all neglect recoil.

An alternative method of simulating atomic fluorescence is to use rate equations, either implemented as a set of differential equations, (Pique \cite{Pique06}), (Hillman \cite{Hillman08}), or by employing Monte Carlo rate equation techniques, (Kibblewhite \cite{KibblewhiteRingberg1}, \cite{KibRep}). In the latter, one follows a more intuitive approach and tracks the time evolution of a single atom, executing photon absorption and emission events according to steady-state cross sections and branching ratios using a random number generator, while accounting for elapsed physical time and recording the atomic state history. After a sufficient time span $\Delta t$ has been simulated, one divides the total number of spontaneously emitted photons by $\Delta t$ to obtain the fluorescent flux. The advantage of this method is that it is easy to understand, and complex physical events such as collisions with different kinds of particles with or without spin exchange can be modeled in a transparent way. Also, the transit of optically pumped atoms across the Doppler spectrum due to velocity-changing collisions and recoil can be easily modeled, hence different velocity classes can be properly coupled, including their atomic polarization exchange (using Bloch equations, such coupling requires the simultaneous solution of sets of $24^2$-dimensional equations per velocity class, which we describe in this work for the first time in application to sodium LGS, to our knowledge).

A serious disadvantage of rate equations is that they rely on steady-state atomic cross sections, limiting the scope to the simulation of events
slow compared to the sodium transition lifetime, (Milonni \cite{Milonni92}), and the combination of the excitation-emission (Rabi) cycling with Larmor precession is hard to implement correctly. Furthermore, when compared to atomic Bloch equations, rate equations neglect the coherences (off-diagonal terms of the density matrix that describe transverse atomic polarization), which is problematic when modeling Larmor precession or linearly polarized light (the latter even for $B\!=\!0$). Finally, Monte Carlo rate-equations need considerable CPU time in order to converge.

There are also approaches that mix aspects of Bloch and rate-equation codes, such as BEACON, which has been adapted to model two-step sodium excitation for polychromatic LGS, (Bellanger \cite{Bellanger}). BEACON neglects atomic collisions and recoil. Guillet de Chatellus (\cite{Chatellus}) reports that it requires on the order of 24\,h per run on a 2.6\,GHz processor, and that the agreement with a pure rate-equation model can be good in certain cases.

Throughout this paper, we highlight what may be called the `three evils' of sodium LGS, ordered by decreasing importance
\begin{enumerate}
\item Larmor precession,
\item Recoil (radiation pressure),
\item Transition saturation (stimulated emission),
\end{enumerate}
and we suggest ways of mitigating them. Larmor precession is powerful enough to thwart optical pumping if the angle between the beam and the field lines is large (Drummond \cite{Drummond}), (Moussaoui \cite{Moussaoui09geomag}). The average 50-kHz redshift that one incurs per spontaneous emission due to atomic recoil can lead to spectral hole burning (depopulation of the respective atoms velocity class) in very bright single-frequency LGS (bandwidth ${}\!<\!10$\,MHz, Hillman (\cite{Hillman08})), but it also offers the opportunity of `snowplowing' the sodium population towards higher velocities in the laser beam direction, requiring continuous blueshifting (chirping) of the laser as first suggested by Bradley (\cite{Bradley}), at the expense of laser complexity. Kibblewhite (\cite{KibblewhiteRingberg2}) is working on an experimental validation of chirping. Ultimately, at high spectral irradiance, stimulated emission becomes relevant, limiting spontaneous emission. Photons from \TB{ stimulated} emission are emitted straight into space, hence becoming useless for LGS\@.

There is some ambiguity in the sodium LGS community about the term `saturation', for it is sometimes used to describe only transition saturation, and sometimes the depopulation of the $F\!=\!2$ upper ground state towards $F\!=\!1$, known as downpumping. Once an atom is in the $F\!=\!1$ state, it can only be excited if the D$_2$b line is pumped as well, which is known as repumping (either by allocating 10--20\,\% of the laser power to the D$_2$b line, or by widening a single laser line to $\geq2$\,GHz). Downpumping in the absence of repumping becomes more severe with increasing laser irradiance and reduces the return flux long before the onset of stimulated emission, hence true transition saturation. Spin-exchange collisions can bidirectionally exchange populations between the ground states and within them (transitions between the ($F\!=\!1,\,m\!=\!\pm1$) and ($F\!=\!2,\,m\!=\!\pm2$) ground substates are particularly strong). Repumping has already been experimentally demonstrated to be able to boost the LGS return flux by a factor of $1.6$, (Denman \cite{Denman06CfAO}, Telle (\cite{Telle08} finds a factor 2.4 through Bloch simulations), using two separate laser beams at different frequencies), and we show in this paper that more than a factor of 3 can be achieved.

To date, the most powerful sodium LGS system is installed at the Starfire Optical Range (SOR) at the Kirtland Air Force Base near Albuquerque, New Mexico, (Denman \cite{Denman06paper}), fed by a single-frequency (10-kHz linewidth) cw laser that emits 50\,W with circular or linear polarization. Due to its location, the median atmospheric seeing at SOR is significantly worse than at most sites of large astronomical telescopes, causing enlarged LGS spots in the mesosphere, and consequently limiting the irradiance. The upcoming generation of civilian 20-W-class LGS in astronomy, such as for the Adaptive Optics Facility of UT4 of the VLT (\cite{AOF}), is therefore expected to venture into unprecedented mesospheric irradiance regimes (much higher laser irradiances have of course been applied to gas cells, albeit at higher gas pressures). Next-generation lasers will emit circularly polarized cw radiation, use repumping, have a linewidth of a few MHz, and are expected to return on the order of $8\times10^6$\,photons/s/m$^2$ on the ground. Quantifying and optimizing their return flux vs.\ the laser parameters is one of the principal purposes of this paper.

So far, we have only focused on photon return flux. What is really desired when designing AO systems, though, is a bright guide star, hence high luminosity concentrated in a small spot size. Compared to the uplink laser irradiance in the sodium layer, the above mentioned saturation effects spatially broaden the LGS return fluorescence distribution by emphasizing the low irradiance regions and dimming the peaks. We can show using physical-optics simulations, Holzl\"ohner (\cite{Holz08Marseille}), that this effect increases the instantaneous spot sizes on a wavefront sensor by about 0.1'', hence it is not negligible. At large angles between the laser beam and the geomagnetic field, downpumping can be mitigated by repumping.

In this work, we present a Bloch-equation method that models \TB{any alkali atom} taking into account spontaneous and stimulated emission, Larmor precession due to the geomagnetic field, arbitrary elliptical light polarization, recoil, on the order of 100 coupled velocity classes with velocity-changing collisions and spin exchange, finite atomic dwell time in the beam (atom replacement), arbitrary laser bandwidth, and repumping. We neglect nonlinear Zeeman shifts and hyperfine coherences since we found them to have a small effect on the result. In contrast to the above cited Bloch simulation publications, we directly compute the steady state solution, which is more efficient than time domain solutions (a single run takes about 2\,s on a modern PC). The program is written in Mathematica and based on the public-domain Atomic Density Matrix package (Rochester \cite{RochesterADM}).

In order to validate our results, we have also implemented a Monte Carlo rate-equation simulation called \emph{Exciter} in Matlab (Holzl\"ohner \cite{HolzCfAO08},\cite{Holz09TRE}), whose results we compare with the Bloch-equation method.

Section~\ref{SecBloch} describes the Bloch-equation method,
Section~\ref{SecParam} gives details about the simulation parameters,
Section~\ref{SecRes} presents the results, suggesting optimal cw sodium LGS
formats, and we conclude in Section~\ref{SecConc}.


\section{Bloch equations}\label{SecBloch}


In order to calculate the observed fluorescence from mesospheric sodium atoms, the evolution of the atoms is modeled using the optical Bloch equations for the atomic density matrix. The density matrix describes the statistical state of an ensemble of atoms in the \TB{state space} of the Na D$_2$ transition. In order to account for atoms with different Doppler shifts, the density matrix is also considered to be a function of atomic velocity along the laser beam propagation direction. (An additional degree of freedom is included to account for laser line broadening as discussed below.) The calculation is semiclassical in the sense that while the atoms are treated quantum mechanically, the light fields are treated classically \TB{(the effect of spontaneous decay must be included phenomenologically since we do not apply field quantization)}. Because the density matrix describes all populations of, and coherences between, the 24 Zeeman sublevels making up the ground and excited states, the calculation describes, in principle, all saturation and mixing effects for essentially arbitrarily large optical and magnetic fields. (In practice, certain coherences in the system are negligible under our experimental conditions and can be neglected in order to increase the computational efficiency.)

In order to perform numerical calculations, the velocity dependence of the density matrix is discretized to describe an appropriate number $n_{v.g.}$ of velocity groups, each with a fixed longitudinal velocity. Because coherences between atoms with different velocities can be neglected, the complete density matrix $\rho$ can be thought of as a collection of $n_{v.g.}$ separate but coupled density matrices, each of dimension $24\times24$.

The evolution of the density matrix is given by a generalization of the Schr\"odinger equation:
\begin{equation}\label{Leq}
    \frac{d}{dt}\rho = \frac{1}{i\hbar}[H,\rho] + \Lambda(\rho) + \beta,
\end{equation}
where $H=H_0+H_E+H_B$ is the total Hamiltonian, with $H_0$ the Hamiltonian for the unperturbed energy structure of the atom, $H_E=-\V{d}\cdot\V{E}$ the Hamiltonian for the interaction of the electric dipole $\V{d}$ of the atom with the electric field $\V{E}$ of the light, $H_B=-\V{$\mu$}\cdot\V{B}$ the Hamiltonian for the interaction of the magnetic moment $\V{$\mu$}$ of the atom with the local magnetic field~$\V{B}$, $\hbar=h/(2\pi)$ where $h=6.626\times10^{-34}$\,Js is Planck's constant, and the square brackets denote the commutator. The term $\Lambda$ in Eq.\,(\ref{Leq}) represents phenomenological terms added to account for relaxation processes not described by the Hamiltonian \TB{(Budker \cite{Budker})}. In our case these relaxation processes include spontaneous decay (omitted from the Hamiltonian due to the semiclassical approximation), collisional spin relaxation (``S-damping'') proportional to $S^2\rho-S\cdot(\rho S)$ (Happer \cite{Happer87}), and the exit of atoms from the light beam due to motion of the atoms and the beam. In addition, there are terms included in $\Lambda$ to describe changes in atomic velocity due to collisions and light-induced recoil, as well as an effective relaxation rate that describes dithering of the laser phase in order to simulate a finite bandwidth. These terms are described in more detail below. \TB{Each relaxation process described by $\Lambda$ includes a corresponding 'repopulation' process, so that the trace over the density matrix for all velocity groups is conserved, corresponding to conservation of the total number of atoms. The repopulation process describing the entrance of atoms into the beam is independent of $\rho$ and so is written as a separate term $\beta$.}

Velocity-changing collisions are treated as hard collisions in which the velocity of the colliding atom is rethermalized in a Maxwellian distribution (no speed memory). The internal state of the atom is assumed to be unchanged.

Light-induced recoil is described phenomenologically by causing a fraction $v_r/\Delta v_{v.g.}$ of the excited-state atoms in each velocity group to be transferred upon decay into the next higher velocity group. Here $v_r$ is the recoil velocity and $\Delta v_{v.g.}$ is the width of the particular velocity group. This model relies on the fact that $v_r=2.9461$\,cm/s (equivalent to a Doppler shift of $50.004$\,kHz) is much smaller than the typical value of
$\Delta v_{v.g.}$.

In order to simulate a finite bandwidth laser, a form of phase dithering is used (frequency or amplitude dithering can also be employed). To avoid resorting to a time-domain calculation, the dithering is implemented in the spatial domain: density matrices are written for two `regions' with light fields that are $\pi$ out of phase with each other, and relaxation terms are included that transfer the atoms between the regions (this doubles the size of the system of equations). The model is that of a laser beam with very fine `speckles' of different phases. The result is an effective laser spectrum of Lorentzian shape with a width proportional to the transfer rate between the regions. This method has been verified by comparison to a time-domain model (implemented for a nuclear-spinless system) in which the light frequency randomly changes with a Lorentzian distribution. Identical results from the two methods are obtained for the case in which the rate that the light frequency changes is faster than the natural
decay rate.

Equation~(\ref{Leq}) supplies a linear system of differential equations for the density matrix elements, known as the optical Bloch equations. Thinking of $\rho$ as a column vector of $n_{v.g}\times24^2$ density matrix elements, the Bloch equations can be written as $\dot\rho=A\rho+b$, where $A$ and $b$ are a matrix and vector, respectively, that are independent of~$\rho$. \TB{The vector $b$ corresponds to $\beta$ and $A$ to the rest of the right-hand side of Eq.\,(\ref{Leq}).}

The laser light field has a frequency component tuned near the D$_2$ $F=2\rightarrow F'$ transition group (D$_{2}$a), and may have an additional `repump' component tuned near the $F=1\rightarrow F'$ transition group (D$_{2}$b). Thus the matrix $A$ has components that oscillate at each of these frequencies. Under the rotating wave approximation (Corney \cite{Corney}), the overall optical frequency is removed from~$A$. However, the beat frequency between the two light-field components remains. This beat frequency can also be removed from the Bloch equations in our case: each frequency component interacts strongly with one transition group and very weakly with the other, so the weak coupling can be neglected for each transition. If, in addition, the small magnetic-field-induced mixing between the two hyperfine ground states is neglected, the beat frequency can be entirely removed from the evolution equations. This makes $A$ time-independent for cw light. To find the steady-state density matrix, we can set $\dot\rho=0$ and solve the linear system $A\rho=-b$. The vectors $\rho$ and $b$ have 322 elements per velocity class (576 if hyperfine states are not neglected), so that the sparse linear equation system has dimension 32,500--65,000 in practice.

To solve the Bloch equations for a particular set of experimental parameters, we first choose an appropriate set of velocity groups. Since the signal is strongly peaked for atoms whose Doppler-shifted resonance frequency is near the light frequency, we can obtain more accurate results for a given number of velocity groups if narrower bins are used for resonant atoms, and wider for off-resonant. We have two methods for doing this.

The first method is to choose two fixed bin sizes, one narrow and one wide, and the number of narrow bins to cluster near each resonance. The wide bins are then used to take up the rest of the Doppler distribution. This method is useful when we don't know beforehand what the spectrum of the signal in velocity space is.

If we have an estimate of the spectrum (obtained using the first method), we can refine it using the second method, which takes advantage of this knowledge. We create a weighting function consisting of three terms: a constant term, which tends to make equal-sized bins, a term proportional to the spectrum, which makes more bins where the signal is large, and a term proportional to the magnitude of the second derivative of the signal, which makes more bins where the signal changes rapidly as a function of velocity. The bin sizes are then found by dividing the integral of the weighting function evenly into the chosen number of bins.

The linear system is solved using the implementation of the iterative BiCGSTAB method (stabilized biconjugate gradient, van der Vorst \cite{Vorst}) built-in to Mathematica. This is a Krylov subspace method in which an initial guess is improved by minimizing the residual over a subspace with dimension much smaller than that of the full system. The rate of convergence of the method is increased by pre-multiplication with a block-diagonal preconditioner (approximate inverse of $A$), obtained by setting all terms that connect density matrix elements from different velocity groups to zero, and then inverting the block for each velocity group.

The fluorescent photon flux per solid angle emitted in a
given direction can be found from the steady-state solution for
$\rho$ as the expectation value of a fluorescence operator
(Corney \cite{Corney}).



\section{Simulation Parameters}\label{SecParam}

  \subsection{Determination of parameters}

   \begin{table}
      \caption[]{Simulation parameters and their \TB{standard nominal} values}
         \label{SimParam}
     $$
         \begin{array}{p{0.55\linewidth}lll}
            \hline
            \noalign{\smallskip}
            Variable name      &  \mathrm{Symbol}  & \mathrm{Standard\ value} \\
            \noalign{\smallskip}
            \hline
            \noalign{\smallskip}
\multicolumn{3}{c}{\bf{Laser\ parameters}}\\
Launched laser power in air	& P			& 20\ \mathrm{W}     \\
Mesospheric laser irradiance	& I			& 46 \ \mathrm{W/m}^2     \\
Central D$_2$ vacuum wavelength	& \lambda		& 589.159\ \mathrm{nm} \\	
Polarization ellipticity angle	& \chi		& \pm\pi/4\ \ \mathrm{(circular)} \\
Laser FWHM linewidth		& \Delta f & 0 \\
Repumping power fraction	& q			& 0.12  \\
Repumping frequency offset 	& \Delta f_{ab} & 1.7178\ \mathrm{GHz} \\
            \noalign{\smallskip}
            \hline
	\noalign{\smallskip}
\multicolumn{3}{c}{\bf{Atomic,\ atmospheric,\ and\ mesospheric\ parameters}}\\
Outer turbulence scale      & L_0   & 25\,\mathrm{m} \\
Geomagnetic field in mesosphere			& B	& 0.228\,\mathrm{G}     \\
One-way transmission at $\lambda$ at zenith	& T_a	& 0.84     \\
Average mesospheric temperature		& T_\mathrm{Na}	& 185\,\mathrm{K}     \\
Sodium centroid altitude (a.s.l.)$^{\ast}$		& H_{\mathrm{Na}}	& 92\,\mathrm{km}     \\
Sodium column density				& C_{\mathrm{Na}}	& 4.0\times\!10^{13}\,\mathrm{m}^{-2} \\
Na beam dwell velocity				& v_{\gamma} & 38\,\mathrm{m/s}     \\
Beam atom exchange rate				& \gamma_\mathrm{ex} 	& 1/(6.0\,\mathrm{ms}) \\
Na--N$_2$ v.c.c.$^{\dagger}$\ cross section		& \sigma_{\mathrm{Na-N}_2}	& 0.71\!\times\!10^{-14}\,\mathrm{cm}^2 \\
Na--O$_2$ v.c.c.$^{\dagger}$\ cross section		& \sigma_{\mathrm{Na-O}_2}	& 0.70\!\times\!10^{-14}\,\mathrm{cm}^2 \\
Weighted v.c.c.$^{\dagger}$ rate						& \gamma_\mathrm{vcc} & 1/(35\,\mu\mathrm{s}) \\
Na--O$_2$ spin exchange cross sect. at $T_\mathrm{Na}$	& \sigma_{\mathrm{Na-O}_2}^S	& 0.50\!\times\!10^{-14}\,\mathrm{cm}^2 \\
Weighted spin-exchange rate at $T_\mathrm{Na}$			& \gamma_S & 1/(490\,\mu\mathrm{s}) \\
            \noalign{\smallskip}
            \hline
	\noalign{\smallskip}
\multicolumn{3}{c}{\bf{Launch\ telescope\ (LT)\ parameters}}\\
Zenith angle					& \zeta	& 30^{\circ}     \\
LT altitude (a.s.l.)$^{\ast}$				& H_{\mathrm{tele}}	& 2650\,\mathrm{m}     \\
LT aperture					& D & 40\,\mathrm{cm}     \\
LT beam radius ($1/e^2$)			& w & 0.36 D = 14.4\,\mathrm{cm}     \\
Launched beam rms wavefront error			& \mathrm{WFE}	& 100\,\mathrm{nm}\approx \lambda/6     \\
Polar angle of $\V{B}$ (laser $\|\ z$)	& \theta	& \pi/2    \\
Azimuth of $\V{B}$ (laser $\|\ z$)	& \phi	& \pi/2   \\
            \noalign{\smallskip}
            \hline
         \end{array}
     $$
\begin{list}{}{}
\item[$^{\ast}$] $\!\!\!$a.s.l. = above sea level;\ \ \ $^{\dagger}$v.c.c. = velocity-changing collision
\end{list}
   \end{table}

Many physical constants and the atomic level diagram of Na have been summarized by Steck (\cite{Steck}) and will not be repeated here. Table~\ref{SimParam} lists further simulation parameters, and we walk through it to explain some quantities. We work in MKS units, except that we show magnetic field strength in Gauss ($1\,\mathrm{G} = 10^{-4}\,T$) and atomic cross sections in squared centimeters.

The launched laser power $P$ equals the laser device output beam power, diminished by optical losses in the beam train and launch telescope (LT). The value of $I = 46\ \mathrm{W\!/m}^2$ is the `50\,\% power in the bucket' irradiance $I_{P/2}$, as explained in the next subsection. We mostly deal with circularly polarized light \TB{(the ellipticity angle $\chi=\arctan(\varepsilon)=\pm\pi/4$ denotes respectively LH/RH circular and $\chi=0$ linear polarization; $\pi/2-2\chi$ is the polar angle in the Poincar\'e sphere, and $\varepsilon$ is the ellipticity, hence the major-minor axes ratio of the polarization ellipse)}. The repumping fraction $q$ is the fraction of the total laser power allocated to the repump beam: the D$_2$a beam power is $(1-q)P$ and the power in the D$_2$b beam (tuned $\Delta f_{ab}$ above the D$_2$a frequency) is~$qP$.

The geomagnetic-field strength $B$ has a strong impact on the return flux. Its value varies considerably over the world and can be computed for different mesospheric altitudes using the International Geomagnetic Reference Field (\cite{IGRF}) model. Cerro Paranal in northern Chile, the location of the VLT and the reference site for this work (24.6$^\circ$S, 70.4$^\circ$W), has $B \approx 0.23$\,G at 92\,km altitude, Mauna Kea (Hawaii) has $B \approx 0.35$\,G, and the Starfire Optical Range (Albuquerque, New Mexico) has $B \approx 0.48$\,G, about twice the field strength at Paranal.

The atmospheric transmission $T_a$ at 589\,nm was measured at Paranal to be $0.89$ in photometric nights, (Patat \cite{Patat}). We use a value that is 5\,\% lower to account for higher aerosol levels. The mesospheric temperature, as well as partial gas densities, can be derived using the MSISE-90 (\cite{MSISE-90}) model.

Our values for the sodium layer centroid altitude $H_\mathrm{Na}$ and column abundance $C_\mathrm{Na}$ are obtained from studies taking place for over 30 years in S\~ao Paulo, (Simonich \cite{Simonich}, Moussaoui \cite{Moussaoui09Na}), a site whose latitude differs only by one degree from the ESO Paranal Observatory. We believe that the sodium layer parameter statistics on the seasonal and daily variations are valid for Paranal.

Atomic collisions have a significant effect on the sodium states and hence on the LGS return flux. Since mesospheric sodium is rarefied (the total mass of global mesospheric sodium is about 600~kg), Na--Na collisions are negligible compared to Na--N$_2$ and Na--O$_2$ collisions. Most of these collisions are binary (collision of two molecules). One important effect of collisions is to change the velocity of the atoms, causing diffusion of optically pumped atoms in velocity space, sometimes called v-damping (Happer \cite{Happer87}). Since the masses of N$_2$ and O$_2$ molecules are comparable to that of Na atoms, one can assume that every collision completely randomizes their velocity. The collision rate of a gas of particle mass $M_1$ with another gas type of particle mass $M_2$ and number density $n_2$ is given by (Wright \cite{Wright})
\begin{equation}\label{gam12}
	\gamma_{12} = n_2 \sigma_{12}
	\sqrt{\frac{8k_B T}{\pi}\left(\frac{1}{M_1}+\frac{1}{M_2}\right)},
\end{equation}
where $k_B = 1.3807\times10^{-23}$\,J/K is Boltzmann's constant and $\sigma_{12}=\pi(r_1+r_2)^2$ is the collisional cross section with the effective particle radii $r_1$,~$r_2$. Measuring these radii for velocity-changing collisions is difficult; here we just assume the Van-der-Waals radii of $r_{\mathrm{Na}}=227$\,pm, $r_{\mathrm{N}_2}=250$\,pm, and $r_{\mathrm{O}_2}=245$\,pm (van den Berg \cite{vandenBerg}), (Fishbane \cite{Fishbane}). The effect of other gas species is negligible. With these numbers, we have, for example, $\sigma_{\mathrm{Na-N}_2}=0.72\times10^{-14}\ \mathrm{cm}^2$ and $\gamma_{\mathrm{Na-N}_2} = 3.98\times10^{-10}\mathrm{cm}^{-3}\mathrm{s}^{-1} \times n_{\mathrm{N}_2} = 1/(62.8\,\mu$s) at the sodium centroid ($n_{\mathrm{N}_2} = 4.0\times10^{13}\ \mathrm{cm}^{-3}$).

The sodium layer has a median FWHM thickness of 11.1\,km and its median centroid lies at $H_{\mathrm{Na}}=92$\,km (Moussaoui \cite{Moussaoui09Na}). The gas pressure decreases exponentially with altitude and the collision rate varies across the layer by about one order of magnitude, as shown by the MSISE-90 (\cite{MSISE-90}) atmospheric model. We compute the mean collision rate of Na with N$_2$ and O$_2$, based on a table of $n_{\mathrm{N}_2}$, $n_{\mathrm{O}_2}$, and $T_\mathrm{Na}$ as functions of altitude, weighted by the assumed Gaussian sodium density distribution. The result, which we will use throughout this work, is $\gamma_{\mathrm{vcc}} = \gamma_{\mathrm{Na-N}_2\mathrm{,O}_2} =1/(35\,\mu$s), which is three times higher than Milonni's (\cite{Milonni99}) assumption of $1/(100\,\mu$s). Since the sodium \TB{abundance and layer thickness and altitude} are highly variable and its profile often deviates significantly from Gaussian, one cannot expect a high accuracy in this parameter.

The other important relaxation mechanism beside v-damping is spin-exchange relaxation or S-damping (Happer \cite{Happer87}, Section~13), in particular for Na--O$_2$ collisions. Spin relaxation time constants have been measured between rubidium and metastable triplet helium (He$^\ast$, Dmitriev \cite{Dmitriev}), as well as between rubidium and H$_2$, O$_2$, and N$_2$, (Nagengast \cite{Nagengast}) and sodium and various gases (Ramsey \cite{Ramsey}), (Kartoshkin \cite{Kartoshkin98}). A major difficulty with such measurements is that the overwhelming contribution to S-damping of Na in the mesosphere is due to collisional spin exchange with O$_2$, however, in gas cells O$_2$ oxidizes Na quickly and hence this particular cross section is hard to determine experimentally. Theoretical calculations of the cross section involve Born-Oppenheimer molecular potential curves of doublet/quartet surfaces for Na--O$_2$, analogous to the singlet/triplet curves for Na--Na and have not yet been carried out to our knowledge (Happer \cite{HapperPriv}).

Kartoshkin (\cite{Kartoshkin09}) estimates $\sigma_{\mathrm{Na-O}_2}^S = 0.5\!\times\!10^{-14}$\,cm$^{2}$ at 185\,K, based on spin-exchange cross section measurements of Na-He$^\ast$ and O$_2$-He. However, only $1/2$ of this cross-section is effective in our case (Dmitriev \cite{Dmitriev}, Eq.\,3)
%
%
Setting $\sigma_{12} = \sigma_{\mathrm{Na-O}_2}^S / 2$ in Eq.\,(\ref{gam12}), we obtain $\gamma_{S} = 1/(680\,\mu$s) at 92\,km altitude, which is close to the value $\gamma_{S} = 1/(640\,\mu$s) that Milonni (\cite{Milonni99}) finds through fitting to experiment. Note, however, that Milonni's initial guess was $\sigma_{\mathrm{Na-O}_2}^S = 1.0\times10^{-14}$\,cm$^{2}$, and he does not apply the scaling factor of~$1/2$. Performing the same sodium density weighted averaging over altitude as above, we obtain $\gamma_{S} = 1/(490\,\mu$s), which will be used throughout this work. We will discuss the sensitivity of the Na return flux to variations in $\gamma_{S}$ and $\gamma_\mathrm{vcc}$ in the following section.



 The rms lateral velocity $v_{\gamma} = d\gamma_\mathrm{ex}$ describes the sodium atom exchange into and out of the beam, where $d \approx 23$\,cm is the median FWHM mesospheric speckle diameter \TB{(see the following section)}, and $\gamma_\mathrm{ex}$ is the atom exchange rate. We assume that $v_{\gamma}$ consists of four components that we sum in quadrature since in general they have uncorrelated directions:
\begin{enumerate}[(a)]
 \item gas diffusion orthogonal to the beam,
 \item mesospheric wind orthogonal to the beam,
 \item beam wander caused by atmospheric turbulence, and
 \item LGS beam slewing due to star tracking.
\end{enumerate}
Each of these contributions can be estimated:
\begin{enumerate}[(a)]

\item The diffusion coefficient of Na in air (mostly N$_2$) can be calculated using the Chapman-Enskog formula (Cussler \cite{Cussler}), yielding $D_\mathrm{Na} = 8.56 \times 10^{4}$\,cm$^2$/s, close to Milonni's assumption of $D_\mathrm{Na} = 1.0 \times 10^{5}$\,cm$^2$/s. The effective lateral diffusion velocity across the laser speckle is hence $\!\sqrt{2/3}D/d = 30$\,m/s, where we apply the factor $\!\sqrt{2/3}$ since we consider only the diffusion orthogonal to the beam. Note that $D_\mathrm{Na}$ scales like $\sqrt{T_\mathrm{Na}}/n_\mathrm{N_2}$. At an altitude of about 104\,km the mean free path of sodium exceeds $d$ and the atomic motion within the beam is no longer diffusive (the mean lateral ballistic velocity is then 370\,m/s).


\item We use the horizontal wind model HWM07 which computes zonal and meridional winds at selected mesospheric altitudes and times (Drob \cite{HWM07}). We find a median horizontal wind speed magnitude at Paranal of 20\,m/s. Nocturnal fluctuations by a factor of 2 are common.

\item Physical-optics simulations show an rms beam wander in the mesosphere of about 20\,cm ($\zeta = 30^{\circ}$), (Holzl\"ohner \cite{Holz08Marseille}), a value supported by analytical approximations (Andrews \cite{Andrews}, Ch.\,12, Eq.\,(51)). The beam wanders on the time scale of $\tau_\mathrm{beam} = d/$(atmospheric wind speed) $\approx$~23\,cm~/~10\,m/s = 23\,ms, leading to a beam wander velocity of 8.7\,m/s.

\item The tracking speed of the laser beam in the mesosphere at $\zeta=30^{\circ}$ is about $(2\pi/24\,\mathrm{h})\times L = 7.6$\,m/s, where $L = X(H_{\mathrm{Na}}-H_{\mathrm{tele}}) \approx X\,90\,\mathrm{km}$ is the line-of-sight distance to the guide star centroid with the geometrical length extension factor (airmass) $X=\sec(\zeta)$, equaling the secant of the zenith angle~$\zeta$.

\end{enumerate}
Summing these four velocities in quadrature yields $v_{\gamma} = 38$\,m/s and
hence $\gamma_\mathrm{ex} = 1/(6.0$\,ms).

\subsection{Mesospheric spot size}

Given a certain launched laser power, the spot size and shape determine the mesospheric irradiance and thus have a strong impact on the return flux, particularly in the presence of strong optical pumping and for small magnetic field polar angles~$\theta$. The instantaneous mesospheric laser spot size has been simulated using physical optics (Holzl\"ohner \cite{Holz08Marseille}). Atmospheric turbulence produces a fast changing \TB{(timescale a few milliseconds)} speckle pattern on the sky \TB{due to diffraction}, where the speckles have a FWHM diameter of about
\begin{equation}\label{eq:d}
	d = \frac{L\lambda}{2w},
\end{equation}
and $w$ is the $1/e^2$ irradiance radius of the laser beam at the projector. The number of speckles and their beam wander are governed by the seeing, and the pattern changes on the time scale of a few milliseconds. Note that the medium/long-term spot size as observed on the ground is significantly larger than $d$, and it does, in contrast to $d$, depend directly on the Fried length~$r_0$ (Fried \cite{Fried}), which is a size scale over which atmospheric phase variations remain below $2\pi$. For a typical $r_0=5$--$25$\,cm, assuming an outer turbulence scale of $L_0\!=\!25$\,m, there is a single strong central speckle and a couple of much weaker satellite speckles. 


The LGS return flux depends in general nonlinearly on the mesospheric laser irradiance $I$, and the return light has to pass the turbulent atmosphere again on the downlink to reach the telescope; thus the irradiance distribution $I(x,y)$, as it would appear on a screen at the mesosphere, cannot be directly observed. To account for the effect of turbulence on the return light and characterize speckle pattern statistically, we collect statistics over many simulated realizations of $I(x,y)$ for different turbulence phase screens. We rewrite the usual optical power integral
\begin{equation}
  P\:(T_a)^{\,X} = \int I\; dA = \int_0^{I_\mathrm{max}} I \frac{dA}{dI}\; dI,
\end{equation}
as an integral over irradiance up to the maximum value $I_\mathrm{max}$, where $A$ is area orthogonal to the beam and $(T_a)^{\,X}$ is the atmospheric transmission along the line of sight (zenith angle $\zeta$). The quantity $dA/dI$ can be thought of as a histogram of mesospheric laser irradiance. We define $I_{P/2}$ by
\begin{equation}
	\frac{P\:(T_a)^{\,X}}{2} = \int_0^{I_{P/2}} I \frac{dA}{dI}\; dI
	= \int_{I_{P/2}}^{I_\mathrm{max}} I \frac{dA}{dI}\; dI,
\end{equation}
representing an irradiance of `50\,\% power in the bucket', which will be convenient to compute laser efficiencies later (note that usually $I_\mathrm{max} \gg 2 I_{P/2}$). In order to model a realistic LGS beam scenario, we assume a $D\!=\!40$\,cm LT, a launched Gaussian beam of $1/e^2$ (in irradiance) radius $w=0.72D/2=14.4$\,cm, two different seeing values of 0.6'' and 1.0'' (site monitor seeing at 500\,nm and zenith), and total wavefront errors (WFE) \TB{of the launched beam} of 70\,nm rms ($\approx \lambda/8$) and 100\,nm rms ($\approx \lambda/6$). We simulate the actual mesospheric irradiance distribution and $I_{P/2}$ using physical optics, and record the equivalent beam diameter FWHM$_\mathrm{eff}$ that a single Gaussian spot of the same mesospheric power and same $I_{P/2}$ would have, as well as the speckle size~$d$, as in Holzl\"ohner (\cite{Holz08Marseille}). Table~\ref{EffBeam} summarizes the results. The mean of $I_{P/2}$ on lines 2 and 4 equals 46\,W/m$^2$, which we will use as the reference irradiance in this work. The median values of $d$ in the last column are somewhat above the analytical prediction of Eq.\,(\ref{eq:d}) of 21.2\,cm, mostly due to the finite WFE\@.

   \begin{table}
      \caption[]{Mesospheric spot parameters for a 40\,cm launch telescope}
         \label{EffBeam}
     $$
         \begin{array}{p{0.25\linewidth}lllll}
            \hline
            \noalign{\smallskip}
            seeing ('')   &  \mathrm{WFE\ (nm)}  & I_{P/2}\ \mathrm{(W/m}^2\mathrm{)} &
\mathrm{FWHM}_\mathrm{eff}\  \mathrm{(cm)} & d\ \mathrm{(cm)}\\
            \noalign{\smallskip}
            \hline
            \noalign{\smallskip}
0.6     & 70  & 79.1  & 30.0 & 22.5 \\  
0.6     & 100 & 59.1  & 34.7 & 24.3 \\  
1.0	& 70  & 43.5  & 40.4 & 23.8 \\  
1.0	& 100 & 33.5  & 46.1 & 26.1 \\  
            \noalign{\smallskip}
            \hline
         \end{array}
     $$
   \end{table}


\subsection{Flux quantities}

In the following, we define five flux quantities used in this article:
\begin{itemize}

\item Our simulations compute the return flux per solid angle $\Psi$ as the number of photons per atom and per unit time spontaneously emitted in the direction of the launch telescope, as observed in the mesosphere, with units of photons/s/sr/atom.

\item In order to compare different laser formats at similar irradiances, one can divide $\Psi$ by the laser irradiance (in W/m$^2$) in the mesosphere to derive the \TB{specific atomic return flux} $\psi\!=\!\Psi/I$ in units of photons/s/sr/atom/(W/m$^2$), where area is measured in the mesosphere orthogonal to the beam. Most of our results are expressed in this quantity.

\item To compute the expected return flux \TB{on a receiver at a distance $L$ from the fluorescing atoms, we begin by integrating $\Psi$ over the receiver area $A$ orthogonal to the beam, subtending the solid angle $A/L^2$ to obtain the quantity~$F_m$}.

\item The photon flux on the detector $\Phi$ (unit photons/s/m$^2$) at the sodium column abundance $C_\mathrm{Na}$ (unit atoms/m$^2$) equals 
\begin{equation}
\Phi = \frac{C_\mathrm{Na}X (T_a)^{\,X} F_m}{L^2}. 
\end{equation} 
The airmass $X$ appears in the numerator because the sodium column along the laser beam scales like the airmass (note that the distance $L$ grows by the same factor). We do not account for laser power depletion with increasing propagation distance in the sodium layer because only about $\sigma \times C_\mathrm{Na} \approx 10^{-15}$\,m$^2 \times 4\!\times\!10^{13}~$m$^{-2}\approx 4$\,\% of the laser photons interact with any sodium atoms under the standard conditions of Table~\ref{SimParam}, the other 96\,\% travel unused into space.

\item In order to compare LGS systems at similar powers but different laser formats, beam spot sizes in the mesosphere, atmospheric transmission, and zenith angle, we can divide $F_m$ by the laser power in the mesosphere $P (T_a)^{\,X}$ to arrive at the figure of merit quantity $s_{ce}$ in units of photons/s/W/(atoms/m$^2$)
\begin{equation}\label{Eq_sce} s_{ce} = \frac{F_m}{P\:(T_a)^{\,X}} = \frac{\Phi L^2}{P\:(T_a)^{\,2X} C_\mathrm{Na}X},
\end{equation}
where $(T_a)^{X}$ appears quadratically in the denominator of the last term, accounting for both up- and downlink. The quantity $s_{ce}$ is hence the photon flux on the ground, corrected for its dependence on sodium centroid height and abundance, airmass, atmospheric transmission, and, to first order, launched laser power (since $s_{ce}$ depends strongly on $theta$, we will only compute it for $\theta=\pi/2$ in this work). We will compare LGS laser beam efficiencies based on~$s_{ce}$. Note that in our definition $s_{ce}$ is not a slope efficiency (which would be proportional to $\partial F_m/\partial P$); see also d'Orgeville (\cite{dOrgeville}).
\end{itemize}

\section{Results}\label{SecRes}
We compute $\psi$ using the method and parameters described above. Using the standard conditions of Table~\ref{SimParam}, we obtain $\psi = 258$\,ph/s/sr/atom/(W/m$^2$). In the following, we first provide some numerical validations including scans of $\psi(I)$, followed by optimization studies of $\psi$ when varying light polarization ellipticity angle~$\chi$, repumping frequency offset~$\Delta f_{ab}$, repumping power fraction $q$, and laser linewidth~$\Delta f$. Unless otherwise noted, the standard conditions of Table~\ref{SimParam} apply.

  \subsection{Some numerical validations}

\begin{figure}
   \centering
      \includegraphics[width=8.8cm]{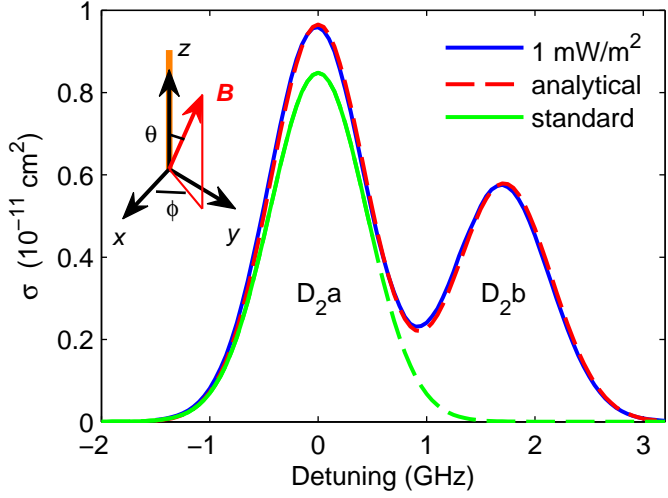}
      \caption{Absorption cross section $\sigma$ vs. detuning from
the Na D$_2$a line center. Solid blue:~simulation
at $I\!=\!0.001$\,W/m$^2$ ($q\!=\!0$), dashed red:~analytical cross section,
green:~simulation of effective cross section at standard conditions ($I\!=\!46$\,W/m$^2$, $q\!=\!0.12$). Inset: reference coordinate system.}
         \label{sigma_vs_detuning}
\end{figure}

Figure~\ref{sigma_vs_detuning} shows a comparison between the analytical Na absorption cross sections given by Eqs.\,(10,11) in (Milonni \cite{Milonni98}) at $T=185$\,K (dashed red line, the sum of two Gaussians) and the simulated effective cross section
\begin{equation}
  \sigma = \frac{h \nu W}{I},
\end{equation}
where $\nu=c/\lambda$ \TB{is the center frequency of the D$_2$a line}, with $c$ the vacuum speed of light, and $W$ is the \TB{actual rate of spontaneous emissions at irradiance $I$} (solid blue curve; the inverse of $W$ is known as the cycle time). \TB{The abscissa shows detuning of the (main) laser line from the D$_2$a line center.} The agreement is excellent. We add the caveat that this agreement is necessary, but not sufficient to prove the validity of our algorithm since effects like Larmor precession, recoil, and stimulated emission have no effect at low irradiance as the Na atom is in thermal equilibrium.

The \TB{steady-state} spontaneous emission rate $W$ cannot exceed $1/(2\tau) = 1/(32.5$\,ns), and hence $\sigma$ tends to zero in the limit of infinite~$I$. At $I=46$\,W/m$^2$ and $q=0.12$ (green curve), the reduction in $\sigma$ is still modest, however. We render the tail of the green curve towards the D$_2$b line center at 1.772\,GHz dashed, where our computational simplification of letting the `D$_2$a laser line' only excite the D$_2$a Na transitions, and analogously allowing the `repumping' line to only excite the D$_2$b transitions, whenever repumping is used ($q\!>\!0$), breaks down. Once the detuning approaches the D$_2$b line center, this assumption obviously becomes invalid. By contrast, the blue curve was computed for $q\!=\!0$ and without using this simplification, and it is valid for any frequency offset.

The inset in Fig.\,\ref{sigma_vs_detuning} sketches the definition of the spherical angles $\theta$ and $\phi$ of the $\V{B}$-vector in a coordinate system where the laser beam is projected along the $z$-axis. The major axis of the polarization ellipse for non-circular polarization is parallel to~$x$.

\begin{figure}
   \centering
      \includegraphics[width=8.8cm]{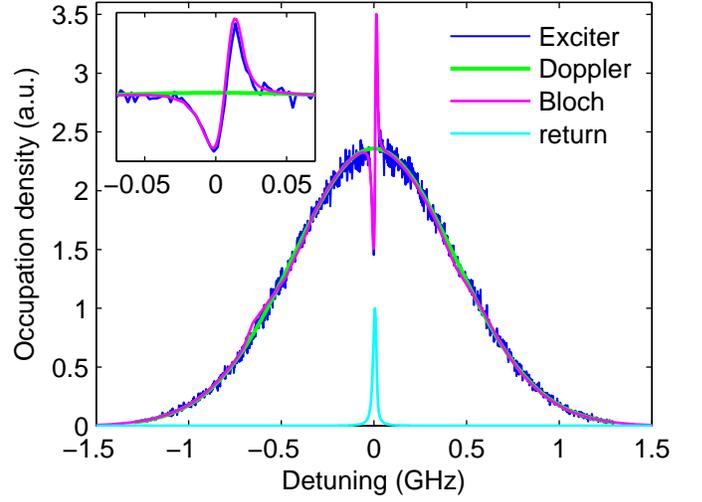}
      \caption{Atomic velocity distribution under standard conditions
(but $\theta\!=\!q\!=\!0$), blue:~Exciter, green:~Doppler profile,
magenta:~Bloch equations, cyan:~return light spectrum \TB{in the atomic frame}. Inset:~zoom on line center.}
\label{occupation}
\end{figure}


Figure~\ref{occupation} shows the simulated atomic velocity distribution under standard conditions, except that Larmor precession and repumping are absent ($\theta\!=\!q\!=\!\gamma_\mathrm{ex}\!=\!0$). \TB{The abscissa shows relative atomic velocity away from the receiver in frequency units (proportionality constant $1/\lambda$)}.
The blue line represents the occupation histogram from the Monte Carlo rate-equation simulation Exciter after sampling $10^6$ different atomic velocities within the Doppler distribution ($8.5\times10^6$ Monte Carlo steps simulating $26.6$\,s of physical time). The green curve is a Gaussian with the theoretical Doppler FWHM width of $1.033$\,GHz at $T_\mathrm{Na}=185$\,K, representing the thermal equilibrium, and the magenta curve shows the occupation probability obtained from the Bloch-equation simulation. Finally, the cyan curve depicts for comparison the simulated return flux spectrum in the atomic frame, \TB{which in the present case of single-frequency excitation is close} to the sodium natural line shape of a Lorentzian with a FWHM of $1/(2\pi\tau)\!=\!9.8$\,MHz (plotted at arbitrary vertical scale).

The effect of spectral hole burning is quite striking; in fact the occupation at the D$_2$a line center is depleted to $64$\,\% below the green Doppler curve. The atomic population, as seen from the telescope, is blue-shifted within about one velocity class. We observe that spectral hole burning due to recoil is in general exacerbated if repumping is applied, presumably due to the larger number of excitations per time. On the other hand, hole burning is mitigated when the laser bandwidth is extended at constant laser power because of the reduced spectral irradiance.

In Exciter, we increment the Doppler shift of the simulated atom by 50\,kHz each time a spontaneous emission occurs, which is correct on average. The agreement between Exciter and the Bloch equations is excellent, giving us confidence that the Gaussian velocity distribution and recoil are properly modeled.

At $I=46$\,W/m$^2$, Exciter simulates a spontaneous emission every 9.5\,$\mu$s on average. Of all emissions, 10.2\,\% are stimulated and 89.8\,\% are spontaneous, and 70.6\,\% of all atomic transitions occur along $(F\!=\!2,m\!=\!2)\leftrightarrow(F'\!=\!3,m'\!=\!3)$, indicating efficient optical pumping. We conclude that besides suffering from recoil, we are also starting to saturate this transition. At $I=100$\,W/m$^2$, 17.9\,\%, and at $I=1000$\,W/m$^2$, 55.2\,\% of all emissions are stimulated, respectively. On average, one spontaneous emission occurs every 5.1\,$\mu$s and 1.1\,$\mu$s for these respective irradiances.

\begin{figure}  
   \centering
      \includegraphics[width=8.8cm]{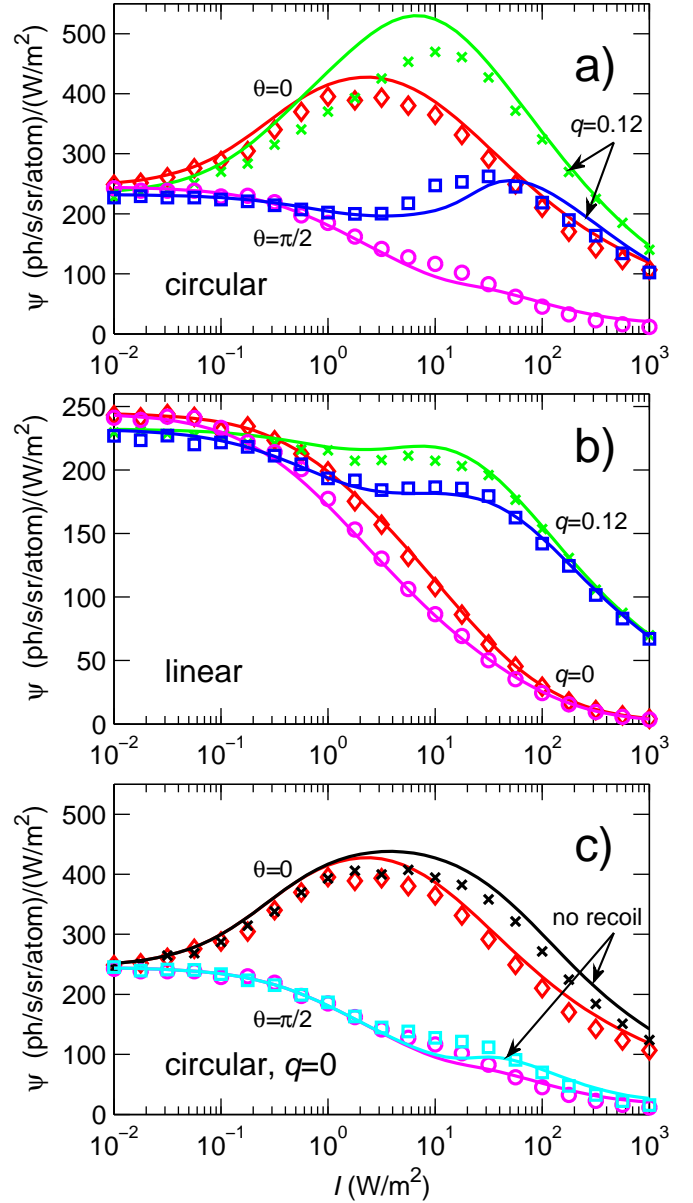}
      \caption{Specific return flux $\psi(I)$.
Lines: Bloch equations, symbols: Monte Carlo rate-equation simulation \emph{Exciter}.
a)~Green, crosses: $\theta=0,\ q=0.12$;
red, diamonds: $\theta=0,\ q=0$;
blue, squares: $\theta=\pi/2,\ q=0.12$ (standard conditions);
magenta, circles: $\theta=\pi/2,\ q=0$
(all with circular polarization and $\gamma_\mathrm{ex}\!=\!0$).
b)~Same as a), but for linear polarization.
c)~Red and magenta lines, diamonds and circles:
as in a); black, crosses and cyan, squares: same, respectively,
without recoil.
}
   \label{psi_vs_I}
\end{figure}

Figure~\ref{psi_vs_I} shows three semilogarithmic plots of $\psi(I)$ for the standard conditions of Table~\ref{SimParam} unless noted otherwise, but neglecting exchange with thermal Na atoms outside the beam ($\gamma_\mathrm{ex}=0$, which is not implemented in Exciter, but causes only a small difference). We hence simulate a single-frequency laser ($\Delta f \rightarrow 0$) tuned to the peak of the D$_2$a line. Solid curves show $\psi$ obtained from the Bloch method, and the symbols indicate the results of Exciter. The green and red curves and symbols represent the case of laser beam parallel to the geomagnetic field ($\theta=0$) with and without repumping, hence in the absence of Larmor precession, respectively. The blue and magenta curves and symbols are the same as green and red, respectively, but with the laser orthogonal to the field ($\theta=\pi/2$). Figure~\ref{psi_vs_I}a shows the case of circular, and Fig.\,\ref{psi_vs_I}b that of linear polarization, the blue and magenta curves for $\theta\!=\!\phi\!=\!\pi/2$ (note the difference in vertical scale).

We can make a number of interesting observations. First, the impact of the magnetic field is profound and reduces the return flux strongly, at some irradiances by several times. At very low irradiance ($I\!=\!10^{-2}$\,W/m$^2$), the atom is in thermal equilibrium and all magnetic sublevels are nearly equally populated. Optical pumping sets in with increasing $I$ if using circular polarization, but Larmor precession is powerful enough to completely suppress it at $\theta=\pi/2$, as evident from the monotonically falling magenta curve. Conversely, in the absence  of Larmor precession $\psi$ strongly peaks near $I\!=\!2$\,W/m$^2$.

The green and blue curves portray the `healing' effect of repumping ($q=0.12$, i.e., 12\,\% of the laser power shifted to D$_2$b). In the absence of Larmor precession, the peak is more pronounced and shifted to higher~$I$ (green). The highest impact occurs when the influence of the magnetic field is highest ($\theta=\pi/2$), where $\psi$ remains shallow until $80$\,W/m$^2$ and then decays gently. This finding refutes the often heard notion in the LGS community that `one should always stay away from the circular polarization saturation irradiation ($I_\mathrm{sat,circular} = 62.6$\,W/m$^2$, Steck (\cite{Steck})),
to avoid saturation'.

At very low $I$, repumping is ineffective and even slightly decreases the return flux due to the smaller cross section of the D$_2$b transition. At high $I$, repumping can readily compensate the degrading effect of Larmor precession; in its absence, $\psi$ decays to zero. For linear polarization, this decay even occurs for $\theta\!=\!0$ (red curve) due to depopulation of the upper ground state. Conversely, circular polarization `rescues' many atoms into the $(S,F\!=\!2,m\!=\!\pm2) \leftrightarrow (P,F'\!=\!3,m'\!=\!\pm3)$ transition, in which they are safe from downpumping.

Note that besides increasing the return flux by several times, a flatter function $\psi(I)$ also leads to smaller observed spot sizes since spatial power-broadening is reduced, a very welcome bonus.

Another interesting observation is that $\psi$ has a local maximum near $I = 80$\,W/m$^2$ for the case of $\theta=\pi/2$ (bump in blue and magenta curves). Here we can observe the effect of a magnetic resonance in which the Rabi cycle time in the most strongly excited velocity classes becomes similar to the Larmor precession time of
\begin{equation}
    \tau_L = \frac{h}{\mu_B |g_F| B},
\end{equation}
where $\mu_B$ is the Bohr magneton and $g_F$ is the Land\'e factor that depends on $F$ ($g_F\!=\!\mp 1/2$ for the $F\!=\!1,\,2$ sodium ground states, respectively). On Paranal, $\tau_L\!\approx\!6.3\,\mu$s \TB{(159\,kHz)} and at SOR we find $\tau_L\!\approx\!3.0\,\mu$s (\TB{333\,kHz}, $B\!=\!0.48$\,G). On resonance, the precessing atoms tend to have the same orientation towards the laser whenever the excitations occur (same $m$), limiting the population mixing and consequently mitigating downpumping. We caution the reader, however, that the location and height of the bump depend on the simulation parameters (e.g.\ the bump almost vanishes for $\gamma_\mathrm{vcc}=1/(100\,\mu$s) or $B=0.48$\,G).

\begin{figure}
   \centering
      \includegraphics[width=8.8cm]{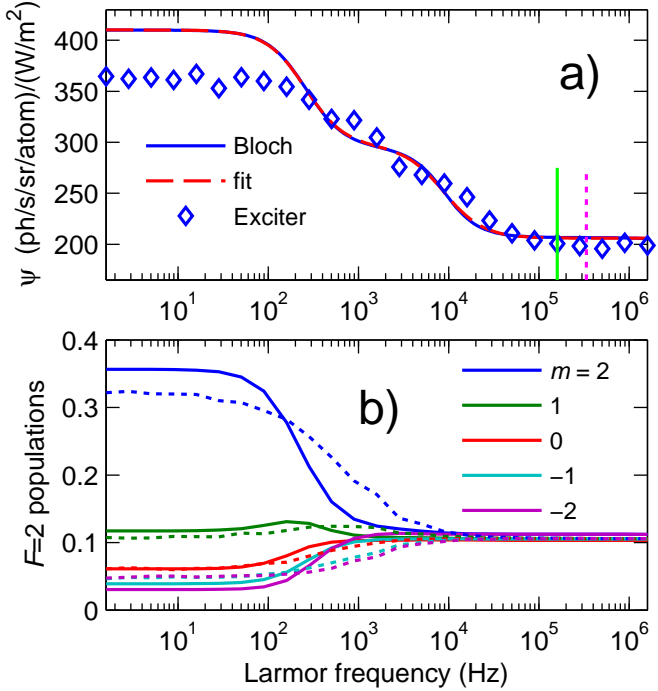}
      \caption{Magnetic field impact.
a)~Blue curve: $\psi$ as a function of the Larmor
frequency $1/\tau_L$ at $I\!=\!1$\,W/m$^2$.
Red dashed curve (almost obscuring the blue curve):~fit
(sum of two Lorentzians). Blue diamonds:~Exciter.
Solid green (dashed magenta) lines: $1/\tau_L$ at Paranal (SOR).
b)~$F\!=\!2$ fractional sublevel populations, solid (dashed) curves:
Bloch (Exciter) simulation.}
   \label{psivsOmegaLarmor}
\end{figure}

The agreement between the Bloch calculation and Exciter is very good in almost all cases, given the completely different nature of the two methods, with the
Bloch code usually yielding the higher values of~$\psi$. Exciter finds the magnetic resonance bump at smaller $I$ due to difficulties with the proper simulation of two concurrent effects on similar, but not equal, time scales in our Monte Carlo rate-equation. Also the agreement between the green curve and the green crosses is somewhat poor around $I=0.2$--20\,W/m$^2$, presumably due to the simplified way in which Exciter models S-damping and/or the absence of coherences.

Figure~\ref{psi_vs_I}c shows the impact of recoil. The red and magenta curves and symbols are as in subplot a), while black and cyan crosses and squares denote the same, respectively, but neglecting recoil. The magnetic resonance is better visible in the cyan than in the magenta curve. Recoil leads to a significant reduction in $\psi$ above 2\,W/m$^2$.

Figure~\ref{psivsOmegaLarmor}a shows $\psi$ as a function of the Larmor frequency $1/\tau_L$ (blue curve) at $I=1$\,W/m$^2$. The red dashed curve, overlapping the solid blue curve, is a fit function composed of the sum of two Lorentzians of different widths, centered at the origin, plus a constant term. The (half-)width of the narrower Lorentzian ($517\,\mathrm{Hz}\!=\!0.25 \gamma_S$) is determined by the S-damping resonance, and that of the broader Lorentzian ($17.9\,\mathrm{kHz}\!=\!0.63 \gamma_\mathrm{vcc}$) by the velocity-changing collision rate, and the widths change proportionally when varying $\gamma_S$ or $\gamma_\mathrm{vcc}$. However, both resonances are somewhat power-broadened. The geomagnetic field is strong enough at $I=1$\,W/m$^2$ to place us on the lowest terrace of the blue curve, as indicated by the vertical lines. The blue diamonds show the result of Exciter for comparison, also exhibiting the terraces.

Figure~\ref{psivsOmegaLarmor}b shows the corresponding relative populations of the five $S,\,F\!=\!2$ upper ground state sublevels ($m\!=\!-2 \ldots 2$), where the solid lines indicate Bloch equations and dotted lines Exciter. For $1/\tau_L\!>\!300$\,Hz, the sublevel populations collapse to the same value due to Larmor-induced sublevel mixing. Conversely, for $1/\tau_L\!<\!300$\,Hz, the populations diverge and the laser pumps the $m=2$ sublevel (blue) most strongly. When increasing the irradiance to $I=46$\,W/m$^2$ (not shown in the plot), the curves in b) would look similar, but the divergence point shifts upward to $\approx 1$\,kHz. Furthermore, when plotting the sublevel populations at $I=46$\,W/m$^2$ as a function of $\theta$, one observes that the magnetic field at Paranal is just strong enough to collapse the populations for $\theta\rightarrow\pi/2$. In other words, by scanning $\theta$ across the sky with a narrow-band 20-W-class laser, we can observe sodium excitation all the way from full optical pumping to no pumping at all, leading to a severe return flux penalty, as shown in the following figures.

   \begin{table}  
      \caption[]{Sensitivity of $\psi$ to a 1\,\% perturbation
in some simulation parameters under standard conditions.}
         \label{SimParamSens}
     $$
         \begin{array}{p{0.55\linewidth}lll}
            \hline
            \noalign{\smallskip}
            Name of perturbed variable  &  \mathrm{Symbol}  & \psi~\mathrm{change\ (\,\%)} \\
            \noalign{\smallskip}
            \hline
            \noalign{\smallskip}
Mesospheric laser irradiance        & I		& \ \ \, 0.025	\\
Geomagnetic field in mesosphere     & B		& -0.33 \\
Average mesospheric temperature     & T_\mathrm{Na}	& -0.48 \\
Beam atom exchange rate			  & \gamma_\mathrm{ex} 	& \ \ \, 0.0025 \\
Weighted v.c.c.$^{\ast}$ rate 				& \gamma_{\mathrm{vcc}} & \ \ \, 0.15 \\
Weighted spin-exchange rate at $T_\mathrm{Na}$	& \gamma_S	& \ \ \, 0.013 \\
Recoil frequency (50\,kHz)		& 		& -0.24	\\
            \hline
         \end{array}
     $$
\begin{list}{}{}
\item[$^{\ast}$] v.c.c. = velocity-changing collision
\end{list}
   \end{table}

Table~\ref{SimParamSens} lists the sensitivity of $\psi$ w.r.t.\ a 1\,\% perturbation in some simulation parameter $p$, more precisely $\psi(1.01p)/\psi(p) - 1$, where all other parameters are those of Table~\ref{SimParam}\@. A value of $b$\,\% in the last table column thus indicates that $\psi(p) \propto p^b$ in some range around the chosen value of $p$. We have selected only those parameters that we will not study in greater detail in the following subsections (except~$I$). In addition, we have excluded those parameters around which $\psi$ is stationary ($\psi^\prime(p)=0$), and also those parameters whose influence on the observed return flux is obviously linear, such as~$C_\mathrm{Na}$.

From Table~\ref{SimParamSens}, we notice that with the parameters of Table~\ref{SimParam} the sensitivity to changes in $I$ is small. A comparison with Fig.\,\ref{psi_vs_I}b (dashed blue curve) shows that $\psi(I)$ is very shallow near $I=46$\,W/m$^2$ (the slope is actually weakly positive due to the magnetic resonance bump), meaning that $\Psi(I)=\psi I \propto I$, i.e.\ the absolute photon return grows linearly with the irradiance. The dependence of $\psi$ on $B$, however, is strong: Switching from $B=0.23$ (Paranal) to $B=0.48$ (SOR, factor $2.11$) decreases $\psi$ by a factor of~$0.76$! The influence of temperature is significant as well since the width of the velocity distribution scales like $T_\mathrm{Na}^{1/2}$, and conversely its normalization (the number of Na atoms per velocity class) scales like $T_\mathrm{Na}^{-1/2}$. However, increasing $T_\mathrm{Na}$ from 185\,K to 200\,K, which may be a typical seasonal variation, reduces $\psi$ by only~$3.7$\,\%, and hence common temperature variations do not directly influence $\psi$ much (we note that temperature influences the sodium abundance). Both $\gamma_\mathrm{ex}$ and $\gamma_S$ have a small influence on~$\psi$, mainly due to the presence of repumping, as Milonni (\cite{Milonni99}) has also noted (for $q=0$, the sensitivities are $0.049$\,\% and $0.56$\,\%, respectively).

We can draw two lessons here: There is little point in spending much effort improving the spin-exchange cross section $\sigma_{\mathrm{Na-O}_2}^S$ estimate, and, secondly, repumping makes $\psi$ more robust towards some parameter variations, leading to higher stability in numerical simulations and possibly also in experiment. The variation of $\psi$ with $\gamma_\mathrm{vcc}$ is much larger than with $\gamma_S$, and at first sight surprisingly, the derivative is positive. We explain the positivity by the fact that collisions mitigate spectral hole burning due to recoil (neglecting recoil, the derivative is $-0.089$\,\%). Finally, we have included the average recoil frequency shift in the list, although it depends only on fundamental constants and $\lambda$ (more precisely $h/(m_\mathrm{Na}\lambda^2) = 50.0$\,kHz, with the atomic mass of Na $m_\mathrm{Na}=3.819\times10^{-26}$\,kg), in order to demonstrate the importance of proper recoil modeling at $I > 1$\,W/m$^2$.

In the following subsection, we will show that the laser parameters of Table~\ref{SimParam} are close to optimal, given the other conditions of the table. We focus on the case $\theta=\pi/2$ where the laser beam is directed orthogonal to the geomagnetic field (the worst case, for which LGS lasers must be designed to achieve a given return flux requirement).

\subsection{Optimization of $\psi$}

  \subsubsection{Polarization ellipticity angle}

   \begin{figure}
   \centering
      \includegraphics[width=8.8cm]{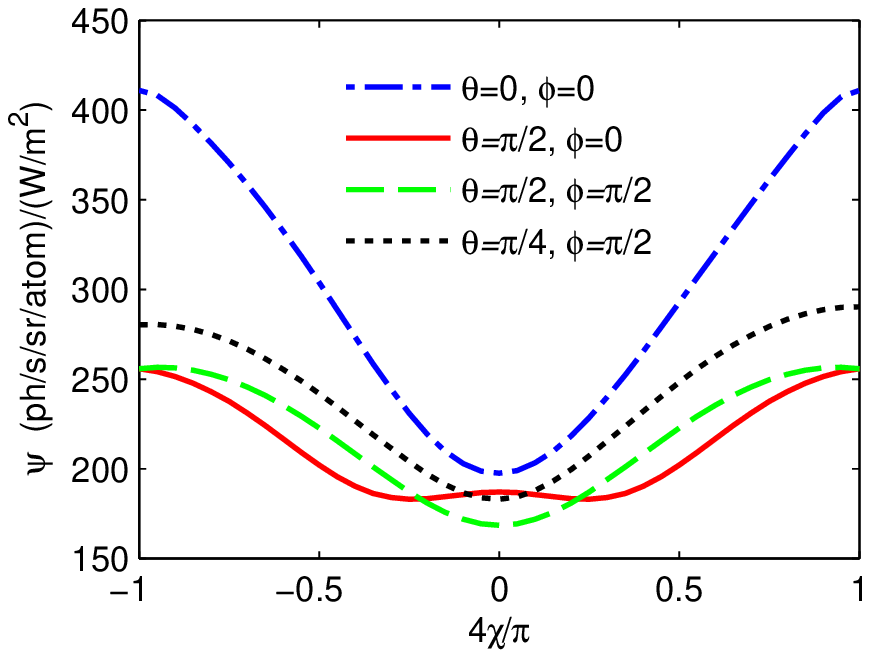}
      \caption{$\psi$ as a function of laser polarization ellipticity angle
$\chi$, where $4\chi/\pi = \pm 1$ (0)
marks circular (linear) states. The curves indicate different
combinations of the magnetic field polar angle $\theta$ and azimuth angle
$\phi$. Note that $\psi(\chi)$ is highest for circular
polarization in all cases.}
         \label{psi_vs_pol}
\vspace{0.5cm}
   \centering
      \includegraphics[width=8.8cm]{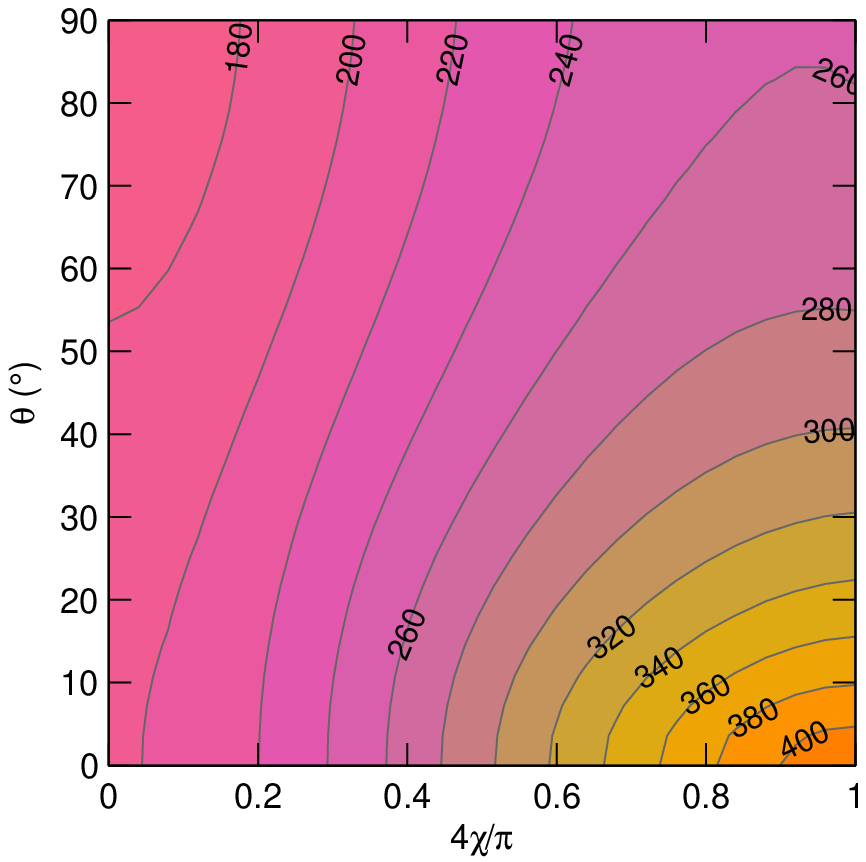}
      \caption{Contour plot of $\psi(\chi,\theta)$ in ph/s/sr/(W/m$^2$) for magnetic azimuth $\phi\!=\!\pi/2$ at $I=46$\,W/m$^2$.}
         \label{psipolThetaBPhiB90}
   \end{figure}

Figure~\ref{psi_vs_pol} plots $\psi$ as a function of the laser polarization ellipticity angle~$\chi$. To our knowledge, this is the first publication to discuss $\psi(\chi)$ for LGS\@. The axis limits $\pm 1$ of the quantity $4\chi/\pi$ denote circular polarization (the poles of the Poincar\'e sphere; in our convention, $4\chi/\pi = +1$ pumps towards increasing $m$), and $\chi=0$ indicates linear polarization. The curves pertain to different combinations of $\theta$ and $\phi$, the polar angle and azimuth of the $\V{B}$ vector in a system where the laser points along $z$ (note that for $\theta\!=\!0$, the azimuth $\phi$ becomes meaningless at any light polarization. Conversely, when using any laser that is not purely circular polarized, $\phi$ \emph{does} influence the return flux, which is often forgotten. When using purely linear polarized light, $\theta\!=\!\pi/2$, $\phi\!=\!0$ is equivalent to setting $B\!=\!0$ within our convention, while $\phi\!=\!\pi/2$ induces the strongest effect from Larmor precession.)

The cosine-like shape of $\psi(\chi)$ presents an initially gradual decrease of the return flux from $4\chi/pi\!=\!\pm 1$, meaning that high polarization purity is not required when pumping the sodium with circularly polarized light. If one is willing to accept a decrease in $\psi$ of 5\,\% along the solid red curve, it is enough to maintain $4|\chi|/\pi \geq 0.8$, i.e., down to an ellipticity angle of $|\chi| = 36^\circ$. The polarization extinction ratio (PER) in dB is given by $\mathrm{PER} = -20\log_{10}\tan|\chi|$ and assumes the values of 0 and $\infty$ for circular and linear polarization, respectively. The condition of $4|\chi|/\pi \geq 0.8$ implies $\mathrm{PER} \leq 6.4$~dB. This insensitivity is good news for the design of launch telescope optics.

Figure~\ref{psipolThetaBPhiB90} shows a contour plot of $\psi(\chi,\theta)$ in ph/s/sr/(W/m$^2$) for $\phi=\pi/2$ (worst case) under standard conditions. Current LGS systems, even SOR's FASOR laser, operate at lower `$P/2$' irradiances than $I_{P/2}=46$\,W/m$^2$ and/or at linear polarization, for which $\psi(\theta)$ varies slowly. At higher $I$ and circular polarization, $\psi(\theta)$ is a steeper function (along the right vertical plot edge). For $\theta > 45^\circ$, also $\psi(\chi)$ varies slowly near $4|\chi|/\pi = 1$, indicating again that the circular polarization purity does not have to be very high.

\subsubsection{Repumping frequency offset}

   \begin{figure}
   \centering
      \includegraphics[width=8.8cm]{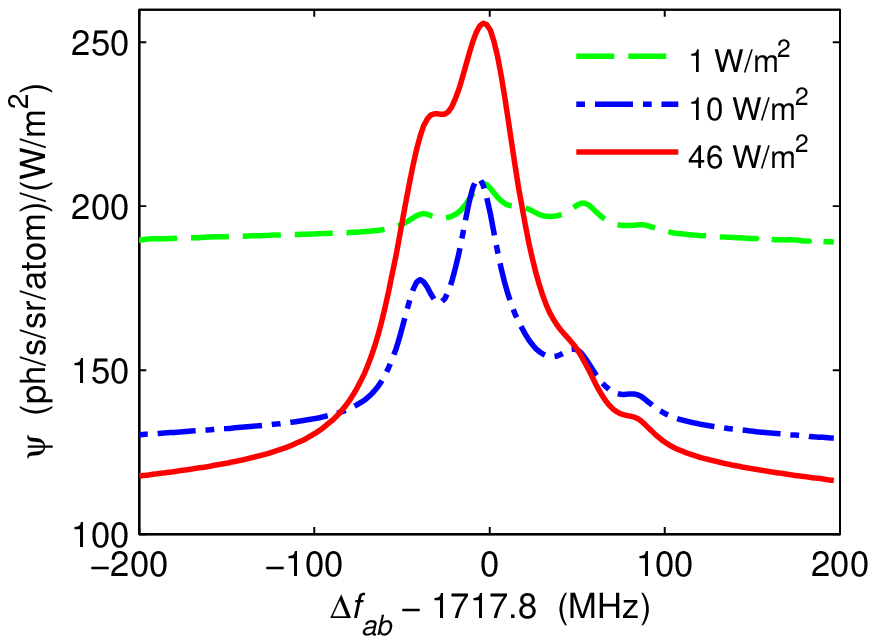}
      \caption{$\psi$ as a function of repumping laser line frequency offset
$\Delta f_{ab}$ at $I = 1$\,W/m$^2$ (dashed green), $I = 10$\,W/m$^2$
(dash-dotted blue), and $I = 46$\,W/m$^2$ (solid red).}
         \label{psi_vs_Delta_f_ab}
\vspace{0.5cm}
   \centering
      \includegraphics[width=8.8cm]{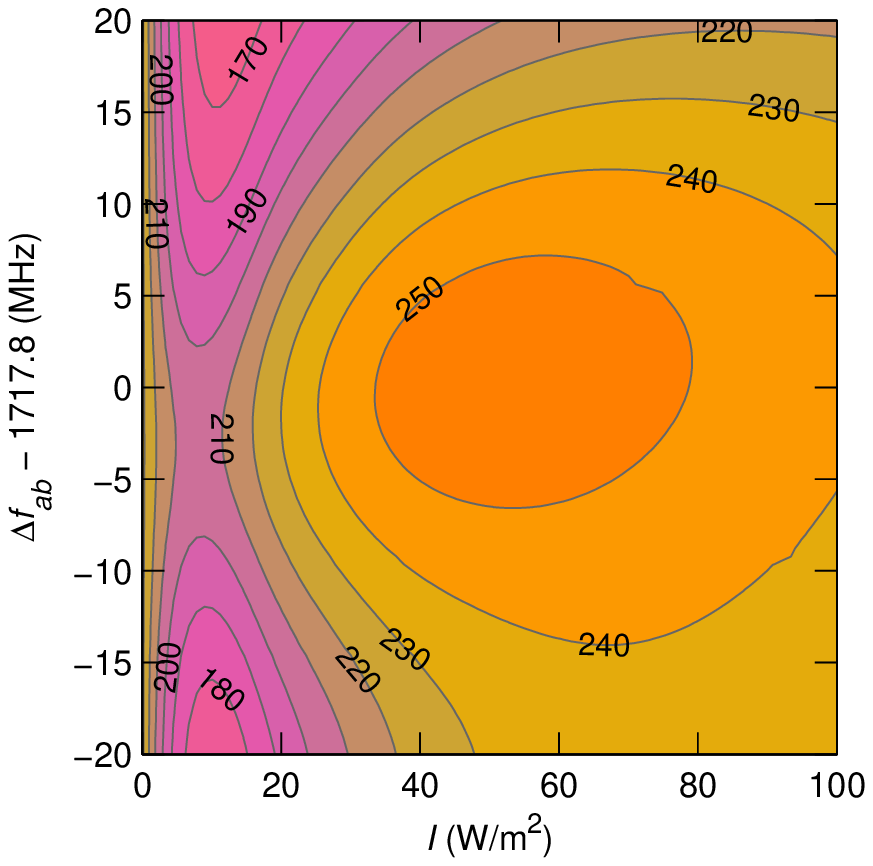}
      \caption{Contour plot of $\psi$ in ph/s/sr/(W/m$^2$)
as a function of $I$ and $\Delta f_{ab}$.}
         \label{psi_Delta_f_ab_IMeso}
   \end{figure}

Figure~\ref{psi_vs_Delta_f_ab} shows $\psi$ as a function of $\Delta f_{ab}$, the repumping frequency offset between D$_2$a and D$_2$b, relative to 1.7178\,GHz, for two single-frequency laser lines and $q=0.12$ at $I=1$\,W/m$^2$ (dashed green), $I = 10$\,W/m$^2$ (dash-dotted blue), and $I = 46$\,W/m$^2$ (solid red). The repumping from $(S,F\!=\!1) \rightarrow (S,F\!=\!2)$ requires excitation to the states $(P,F'\!=\!1)$ or $(P,F'\!=\!2)$ since no transitions $(S,F\!=\!1) \rightarrow (P,F'\!=\!3)$ are allowed.

Figure~\ref{psi_Delta_f_ab_IMeso} is a contour plot of $\psi(I,\Delta f_{ab})$, zooming into the region of $\Delta f_{ab}\!=\!1717.8 \pm 20$\,MHz. In order to harness the full improvement of repumping, one needs to adjust $\Delta f_{ab}$ with a precision of a few MHz, particularly when at intermediate irradiance ($I\!\approx\!10$\,W/m$^2$). At larger irradiance, power-broadening washes out the peak (a laser beam has irradiances $0\leq I\leq I_\mathrm{max}$ in the mesosphere, located along a horizontal line in the diagram). The tolerance in $\Delta f_{ab}$ is so narrow because repumping is most efficient if a given Na atom that has been downpumped to the $F\!=\!1$ lower ground state can be reexcited before its velocity changes due to a collision, for which the frequency offset of the repumping laser line should lie between the $(S,F\!=\!1) \rightarrow (P,F'\!=\!1,2)$ transition frequencies minus the $(S,F\!=\!2) \rightarrow (P,F'\!=\!3)$ transition frequency that the main laser line is tuned to, i.e., close to $1717.8$\,MHz.

  \subsubsection{Repumping power fraction}

   \begin{figure}
   \centering
      \includegraphics[width=8.8cm]{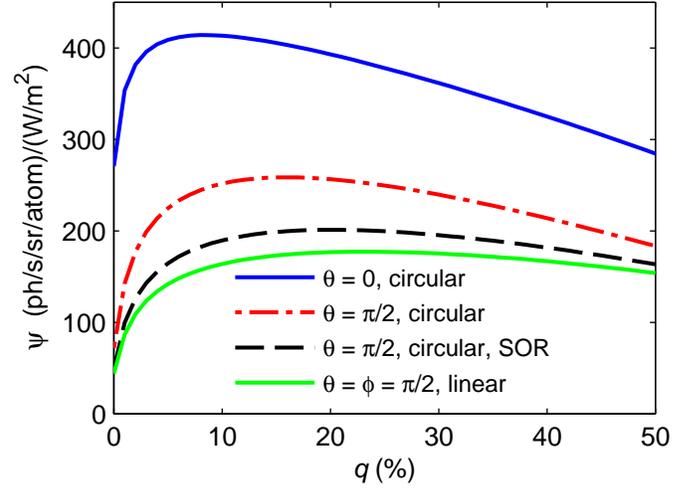}
      \caption{$\psi$ as a function of repumping power fraction $q$ for
different field angles~$\theta$. Dash-dotted red curve:~standard conditions,
solid blue: $\theta=0$ (circular polarization, no Larmor precession),
dashed black:~standard conditions, but $B=0.48$\,G (SOR),
solid green:~linear polarization ($\theta=\phi=\pi/2$).}
         \label{psi_vs_q}
   \end{figure}

   \begin{figure*}
   \centering %
   \includegraphics[width=18cm]{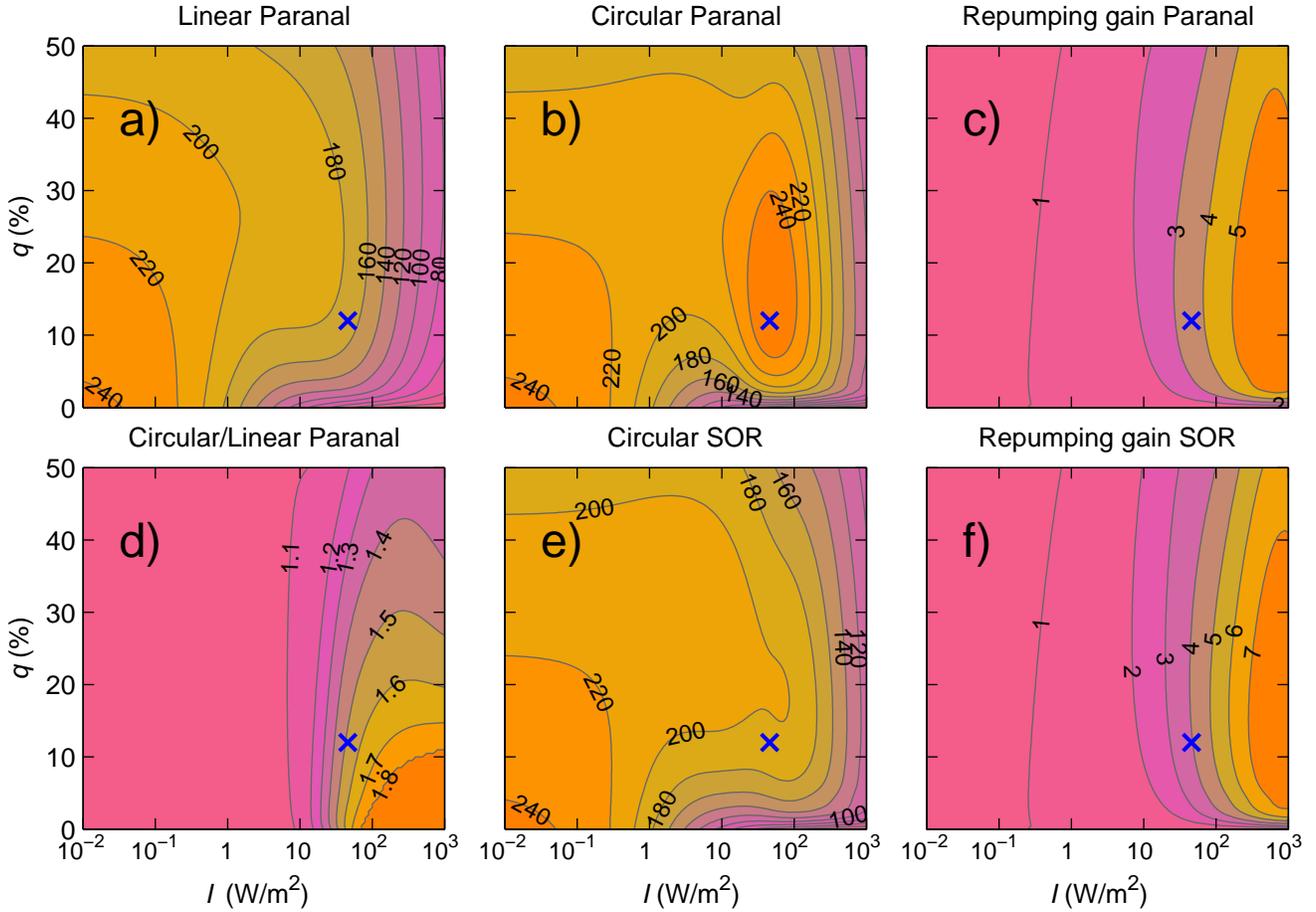}
   \caption{
a,b)~Contour plots of $\psi(I,q)$ in ph/s/sr/(W/m$^2$) at Paranal for linear and circular polarization, respectively;
d)~The ratio $\psi(\mathrm{circular})/\psi(\mathrm{linear})$ at Paranal;
e)~Same as b), but for SOR geomagnetic field;
c,f)~Repumping gain $\psi(I,q)/\psi(I,q\!=\!0)$ at Paranal and SOR, respectively,
circular polarization. Blue crosses: $I=46$\,W/m$^2$, $q=0.12$.}
\label{psi_I_q}
    \end{figure*}

Figure~\ref{psi_vs_q} portrays the calculated $\psi(q)$ for different magnetic field polar angles~$\theta$ and field strengths~$B$. The dash-dotted red curve is for standard conditions of Table~\ref{SimParam}, the solid blue curve for $\theta=0$ (i.e., circular polarization and no Larmor precession), the dashed black curve is for standard conditions, but with the magnetic field at SOR, and the solid green curve is for linear polarization (worst case $\theta\!=\!\phi\!=\!\pi/2$).

All curves peak around $q=8$--$25$\,\%, and the maximum occurs at larger $q$ the stronger the magnetic field effect is. Note that $\psi(q)$ peaks at $q\!>\!0$ even in the absence of Larmor precession (blue curve) since there is always a chance that atoms decay to the lower ground state. Moreover, $\psi(q)$ is very steep at $q=0$, rewarding even the weakest repumping, which is the reason why broadline lasers (${}\!\!\approx\!2$\,GHz) often do reasonably well, despite their poor spectral overlap with $\sigma(\nu)$ (see Fig.\,\ref{sigma_vs_detuning}). Our simulations, however, indicate that a narrow-band laser with 12\,\% repumping beats the return flux of a single-line 2\,GHz-linewidth laser by a factor of 3.7 under standard conditions (see next subsection).

For circular polarization we find $\psi(q\!=\!0.12)/\psi(q\!=\!0) \approx 3.5$ at $I=46$\,W/m$^2$, underlining again that \emph{repumping is extremely beneficial, in particular in future LGS systems involving high irradiance.}


Figure~\ref{psi_I_q} shows six semilogarithmic contour plots of $\psi(I,q)$ (note the different color scales of the plots).
The blue crosses indicate the point ($I\!=\!46$\,W/m$^2$, $q\!=\!0.12$).
Plot~a) exhibits $\psi$ for linear polarization at Paranal for $\theta\!=\!\phi\!=\!\pi/2$,
which only varies weakly with~$B$. At the blue cross, $\psi=169$\,ph/s/sr/(W/m$^2$)
(for $q\!=\!0$, it is only 44\,ph/s/sr/(W/m$^2$)!). Plot~b) shows the same for circular
polarization ($\psi=258$\,ph/s/sr/(W/m$^2$) at the blue cross, which is close to the
maximum of $\psi$ in Plot~b) and lies within the magnetic resonance `island').
Plot~d) shows the ratio
$\psi(I,q,4\chi/\pi\!=\!1)/\psi(I,q,\chi\!=\!0)$. Circular
polarization at $\theta=\pi/2$ only improves the return flux at Paranal
if $I\!\geq\!10$\,W/m$^2$ (at smaller $\theta$, there is a much higher gain,
as indicated by Fig.\,\ref{psi_vs_I}). The improvement at the blue cross is 1.52 (with the $B$-field at SOR, it would be only 1.2, but can reach 2.0 for smaller $\theta$).
We have limited the plot range in the lower right corner of d) to 1.8 for better
rendering; in reality, the value exceeds~3.
Plot~e) shows the same as b), but for the $B$-field
of SOR\@. As noted in Fig.\,\ref{psi_vs_q}, the optimal $q$ lies closer to 19\,\%
under this condition. Note that even at $q\!=\!0.5$ (equal power in
D$_2$a and D$_2$b), $\psi\approx 164$\,ph/s/sr/(W/m$^2$) is not much
below its optimum of 201.
Plot~c) depicts the gain in $\psi$ when turning on repumping at standard conditions, i.e., $\psi(I,q)/\psi(I,q\!=\!0)$. Plot~d) shows the same for SOR\@. At the blue crosses $\psi(q\!=\!0.12)/\psi(q=0)$
equals $3.6$ and $4$, respectively.

  \subsubsection{Laser linewidth}

\begin{figure}
\centering
\includegraphics[width=8.8cm]{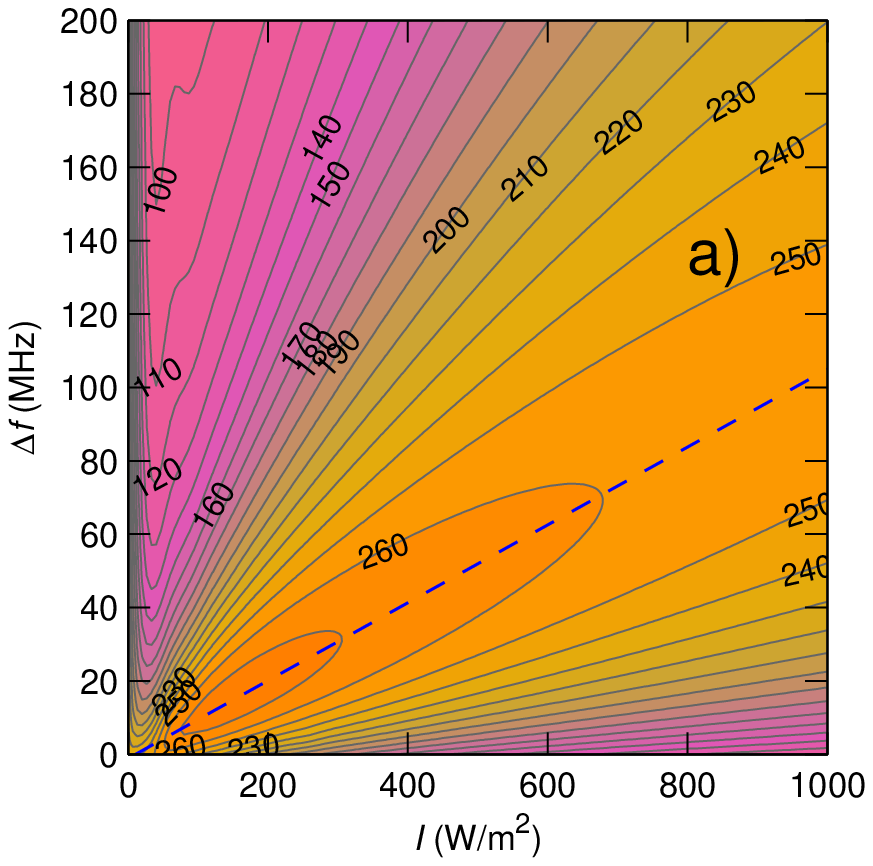}
\vspace{0.5cm}
\centering
\includegraphics[width=8.8cm]{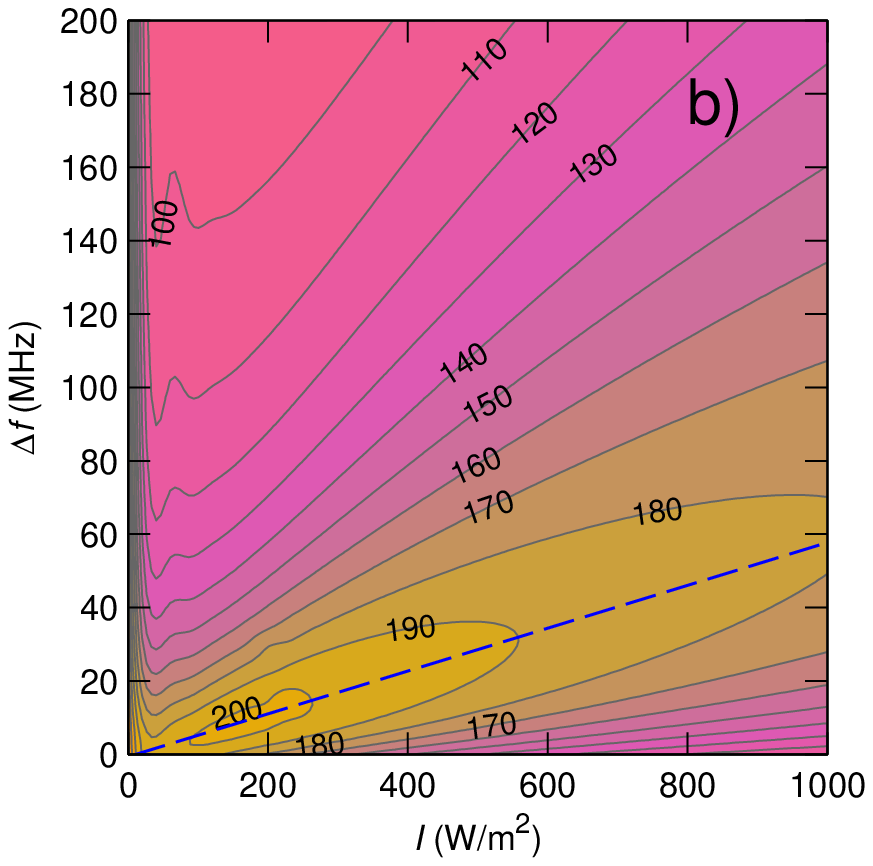}
  \caption{a)~Contour plot of $\psi(I,\Delta f)$ in ph/s/sr/(W/m$^2$)
for standard conditions (Paranal).
b)~Same, but the geomagnetic field of SOR\@. The
optimal spectral irradiance is near 10 and 17\,W/m$^2$/MHz, respectively
(dashed blue lines).}
     \label{psiBwIMeso}
\end{figure}

\TB{So far, we have only simulated single-frequency line lasers. We now turn to lasers with Lorentzian lines of finite bandwidth.} Figure~\ref{psiBwIMeso} shows two contour plots of $\psi(I,\Delta f)$ for conditions at Paranal and SOR\@. The specific return is highest near the dashed blue lines, hence near 10\,W/m$^2$/MHz, or 100\,W/m$^2$ per natural linewidth of sodium of $9.8$\,MHz for Paranal and near 17\,W/m$^2$/MHz for SOR\@. For irradiances below about 20\,W/m$^2$, this linearity does not hold anymore and $\Delta f \rightarrow 0$ will yield higher return flux. Conversely, at $I\!\gg\!1000$W/m$^2$, $\Delta f$ is no longer negligible compared to the Doppler width, and hence the absorption cross section falls off in the wings of the laser spectrum (our simulations for $q\!>\!0$ lose accuracy for wide bandwidths, see the discussion of Fig.\,\ref{sigma_vs_detuning}). In the upper left region of the plot, $\psi$ decreases due to low pumping \TB{that is too weak to overcome the population equalization from Larmor precession}, and in the lower right we suffer from saturation. We note that d'Orgeville (\cite{dOrgeville}) has conducted a related study (d'Orgeville's Fig.\,8 differs from ours in that its horizontal axis shows laser power and not irradiance, and it is rendered in double logarithmic scale).


Based on the dashed blue lines in Fig.\,\ref{psiBwIMeso}, the optimum FWHM laser bandwidth can be estimated by $\Delta f_\mathrm{opt} \approx (0.025\ I_{P/2} - 0.5)/B$, where $\Delta f_\mathrm{opt}$ is in MHz, $I_{P/2}$ in W/m$^2$ with $20\!\leq\!I_{P/2}\!\leq\!1000$\,W/m$^2$, and $B$ in Gauss. We expect this optimal spectral power density to only depend on the relaxation time scales. This scaling relationship is good news for future LGS systems, since it means that we can achieve high return flux even at very large irradiances, as long as we keep increasing the laser bandwidth. Lasers with microsecond pulses designed for mesospheric spot tracking or lidar with duty cycles of 1:20\,--\,1:100, and hence high peak powers, may take advantage of this possibility.

  \subsection{Optimization of laser beams}

In this subsection, we extend the analysis to entire laser beams by
carrying out weighted integrals of $\psi(I)$ for (Gaussian) spot
profiles, from which we compute $s_{ce}$ (Eq.\,(\ref{Eq_sce})).

To begin, we conduct a comparison with the beam return flux computed by Milonni (\cite{Milonni99}). We apply our Bloch code to Milonni's Na conditions of $B\!=\!0.5$\,G, $q\!=\!0$, $\theta\!=\!30^\circ$, $\zeta=0^\circ$, $P\!=\!1$\,W, $2\times I_{P/2}\!=\!0.52$\,W/m$^2$, circular polarization ($4|\chi|/\pi\!=\!1$), no recoil, $\gamma_\mathrm{ex}\!=\!0$, $T_\mathrm{Na}\!=\!200$\,K, $T_a\!=\!1.0$, $\gamma_\mathrm{vcc}\!=\!0$, $\gamma_{S}\!=\!1/(100\,\mu$s)) and obtain $s_{ce} = 335$\,ph/s/W/(atoms/m$^2$), in contrast to Milonni's value $s_{ce} = 240$\,ph/s/W/(atoms/m$^2$) from Eq.\,(30) of that paper. If we switch to $\gamma_{S} = 1/(640\,\mu$s) (long-dashed curve in Milonni's Fig.\,6a), we obtain $s_{ce} = 359$\,ph/s/W/(atoms/m$^2$). For the combination $\gamma_\mathrm{vcc} = 1/(100\,\mu$s), $\gamma_{S} = 1/(640\,\mu$s), we find $s_{ce} = 303$\,ph/s/W/(atoms/m$^2$), underlining the importance of modeling diffusion in atom velocity space. We suspect that the discrepancy in the flux results is partly due to Milonni's (\cite{Milonni99}) greatly simplified S-damping formula Eq.\,(18), which tends to overestimate spin relaxation.

\begin{figure}
\centering
\includegraphics[width=8.8cm]{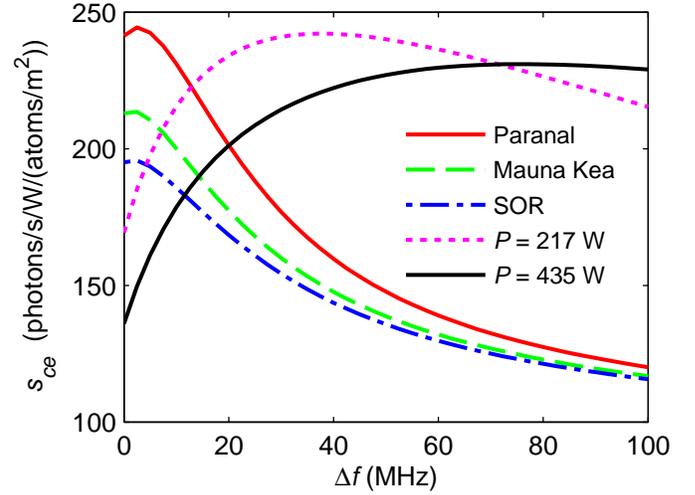}
  \caption{Beam efficiency $s_{ce}$ as a function of FWHM laser
bandwidth $(\Delta f)$ for a $P=20$\,W laser at Paranal (solid red curve, standard conditions), Mauna Kea (dashed green), and SOR (dash-dotted blue),
depending on the local geomagnetic-field strengths
($P\!=\!20$\,W, $\theta\!=\!\pi/2$).
Dotted magenta (solid black):~Laser
at Paranal for $P\!=\!217$\,W ($P\!=\!435$\,W).}
     \label{sce_vs_bw}
\end{figure}

Figure~\ref{sce_vs_bw} shows a plot of the beam efficiency~$s_{ce}$
for different geomagnetic field strengths at $\zeta\!=\!30^\circ$
(solid red:~Paranal, dashed green:~Mauna Kea, dash-dotted blue:~SOR).
The dotted magenta and solid black lines show $s_{ce}$ for Paranal at $P=217$\,W ($I_{P/2} = 500$\,W/m$^2$) and $P=435$\,W ($I_{P/2}\!=\!1000$\,W/m$^2$), respectively. At these irradiances, the $s_{ce}(\Delta f)$ peaks at 38\,MHz and 75\,MHz, respectively, without losing much of its magnitude compared to $P=20$\,W. We expect that pulsed lasers with pulse durations much longer than 16\,ns can enjoy this high specific return. Note that the numerical similarity of $s_{ce}\!\leq\!245$\,ph/s/W/(atoms/m$^2$) with typical values of $\psi$ in ph/s/sr/(W/m$^2$) is a mere coincidence, and the two quantities should not be confused.

We add the caveat that the laser bandwidth is defined here as the \emph{short-term bandwidth}, as opposed to the apparent bandwidth obtained from a long-term spectral measurement of a laser whose central frequency is slowly drifting, or being chirped.



\section{Conclusions}\label{SecConc}

In this article, we describe a modeling method to accurately determine the photon return flux from sodium LGS using Bloch (density matrix) equations. Numerical simulations are crucial, since it is not easy to achieve the desired optimization experimentally.

We summarize some optimization conclusions for cw sodium LGS lasers as follows:
   \begin{enumerate}
	\item The next generation of 20\,W-class (launched) cw laser sodium LGS will achieve unprecedented mesospheric irradiances near $I_{P/2}\!=\!50$\,W/m$^2$ ($I_{P/2}\!\!=\!\!\!{}$ `50\,\% power in the bucket irradiance') in good seeing conditions, using launch telescopes with 30--50\,cm clear aperture diameter.

	\item Atomic Larmor precession due to the geomagnetic field $B$ can
completely suppress optical pumping and thus strongly reduce the LGS return flux, depending on the orientation between the laser beam and the field lines. LGS systems must be dimensioned for the beam pointing that yields the lowest return, i.e., laser beam orthogonal to~$\V{B}$.

	\item In the future LGS irradiance regime, circular polarization
will significantly elevate the return flux, with a rather relaxed polarization extinction ratio of $\mathrm{PER}\leq 6.4$\,dB. The polarization chirality (sense) does not matter.

	\item Repumping (excitation of the D$_2$b line) with
10--20\,\% of the laser power in combination with circular polarization
increases the return flux further by a factor of 3--4\@. Repumping can
moreover alleviate spatial power-broadening and hence reduce the observed
LGS spot size.

	\item The optimum FWHM laser bandwidth can be estimated for a Lorentzian line shape by
$\Delta f_\mathrm{opt} \approx (0.025\ I_{P/2} - 0.5)/B$, where
$\Delta f_\mathrm{opt}$ is in MHz, $I_{P/2}$ in W/m$^2$
($20\!\leq\!I_{P/2}\!\leq\!1000$\,W/m$^2$), and $B$ in Gauss.
As a rule of thumb, the laser bandwidth in MHz should approximately equal
the launched laser power in Watt divided by six, assuming a diffraction-limited
spot size in good seeing. Employing lasers with a single line of 2--3\,GHz
bandwidth is usually several times less efficient.

	\item A laser with the above properties can achieve a specific
return of $s_\mathrm{ce} = 200\!-\!250$\,photons/s/W/(atoms/m$^2$) at a low-$B$ location such as in northern Chile and $s_\mathrm{ce} = 150\!-\!200$\,photons/s/W/(atoms/m$^2$) at a high-$B$ location such as New Mexico (all values for laser beam orthogonal to $\V{B}$).

   \end{enumerate}
\TB{We expect a numerical accuracy of $\pm10$\,\% in our results of~$\psi$ for a given parameter set. However, many physical simulation parameters such as spin relaxation and velocity-changing collision cross sections are still to be measured more accurately (at mesospheric temperatures). The corresponding relaxation rates depend on gas pressure and temperature and thus vary exponentially with altitude. In practice, the absolute return flux uncertainty is completely dominated by sodium abundance and altitude distribution fluctuations (for instance, raising the altitude by 6\,km causes a reduction of $\gamma_\mathrm{vcc}$ from $1/(35\,\mu$s) to $1/(100\,\mu$s), diminishing $s_\mathrm{ce}$  from 245 to 205\,photons/s/W/(atoms/m$^2$)). The accurate return flux computation of a realistic LGS system requires averaging along the beam, accounting for altitude-dependent temperature, gas concentrations, and laser irradiance. Despite these caveats, we do not expect the qualitative conclusions of this work to change when the parameter values are made more precise.}

We also compare the Bloch simulation results with our Monte Carlo rate-equation simulation \emph{Exciter} and find good agreement. However, such Monte Carlo rate-equation methods are based on numerous assumptions, and they always need to be validated against more rigorous methods such as a Bloch equation calculation. Moreover, our Bloch code is orders of magnitude faster than Exciter, with the runtime advantage quickly growing with the desired level of accuracy. Our Bloch simulation code can simulate \TB{any alkali} and is publicly available on our websites (\cite{ESO_Web} and Rochester \cite{RochesterADM}, \TB{requires Mathematica v.\,6 or higher}).

In this work, we treat only steady-state sodium excitation with cw lasers. Pulsed lasers, with pulse durations close to or shorter than the sodium lifetime of 16\,ns, will be dealt with in a forthcoming publication. However, we surmise that our present results are valid for microsecond laser pulses that are useful for mesospheric spot tracking and lidar. On-sky experiments to validate our results with a mobile 10\,W-class laser unit are being prepared at ESO at the moment, and are expected to yield results in 2010\@.

In conclusion, our simulations indicate that we have only begun to realize the full capability of sodium LGS, to be harvested by upcoming generations of laser-assisted AO, both with cw and pulsed lasers. The future of sodium LGS, so to speak, looks very bright.

\begin{acknowledgements}
We thank Craig Denman, Paul Hillman, Peter Hoffmann, Victor Kartoshkin, Ed Kibblewhite, Peter Milonni, Steve Morgan, Brian Patton, John Telle, Yan Feng, Nourredine Moussaoui and Luke Taylor for their valuable input. We would like to give special credit to Will Happer, the father of sodium LGS, for his generous advice. This work was supported in part by the NGA NURI program.
\end{acknowledgements}




\begin{thebibliography}{}

\bibitem[2005]{Andrews}
Andrews, L. C. and Phillips, R. L., 2005,
Laser Beam Propagation through Random Media (SPIE Press, Bellingham),
2nd~ed.

\bibitem[AOF]{AOF}
Adaptive Optics Facility (AOF) for the VLT, European Southern Observatory (ESO),
\href{http://www.eso.org/projects/aot/DSM/}{http://www.eso.org/projects/aot/DSM/}

\bibitem[2004]{Bellanger}
Bellanger, V., Courcelle, A., and Petit, A.,
``A program to compute the two-step excitation of mesospheric sodium atoms for the Polychromatic Laser Guide Star Project'', Comp.\ Phys.\ Comm. {\bf 162}, pp.~143--150 (2004)

\bibitem[online]{vandenBerg}
van den Berg, atomic radius list online at
\href{http://www.ccdc.cam.ac.uk/products/csd/radii/}{http://www.ccdc.cam.ac.uk/products/csd/radii/}

\bibitem[1992]{Bradley}
Bradley, L. C., ``Pulse-train excitation of sodium for use as a
synthetic beacon,'' J.\ Opt.\ Soc.\ Am.\ B {\bf 9}, pp.~1931--1944 (1992)

\bibitem[2002]{Budker}
Budker, D., Gawlik, W., Kimball, D. F., Rochester, S. M., Yashchuk, V. V., and Weis, A., ``Resonant nonlinear magneto-optical effects in atoms,'' Rev.\ Mod.\ Phys.\ \textbf{74}, pp.~1153 (2002)

\bibitem[2008]{Chatellus}
Guillet de Chatellus, H., Pique, J.-P., and Moldovan, I. C.,
``Return flux budget of polychromatic laser guide stars'',
J.\ Opt.\ Soc.\ Am.\ A~{\bf 25}, pp.~400--415 (2008)

\bibitem[1985]{McClelland}
McClelland, J. J., and Kelley, M. H.,
``Detailed look at aspects of optical pumping in sodium'',
Phys.\ Rev.\ A {\bf 31}, pp.~3704--3710 (1985)

\bibitem[1977]{Corney}
Corney, A., 1977, Atomic and Laser Spectroscopy, Oxford
University Press

\bibitem[1997]{Cussler}
Cussler, E. L., ``Diffusion. Mass Transfer in Fluid Systems'', 2nd edition,
Cambridge University Press, 1997

\bibitem[2006a]{Denman06paper}
Denman, C. A., Drummond, J. D., Eickhoff M. L., Fugate, R. Q.,
Hillman, P. D., Novotny S. J., and Telle J. M.,
``Characteristics of sodium guidestars created by the
50-watt FASOR and first closed-loop AO results at the
Starfire Optical Range'', Proc.\ SPIE {\bf 6272}, 62721L (2006)

\bibitem[2006b]{Denman06CfAO}
Denman, C. A. \emph{et~al.}, ``Two-Frequency Sodium Guidestar Excitation
at the Starfire Optical Range'', talk at Center for Adaptive Optics,
Fall Retreat 2006, Fish Camp, California (2006)

\bibitem[2008]{Dmitriev}
Dmitriev, S. P., Dovator, N.A. and
Kartoshkin, V. A., ``Spin Exchange Rate Constant for Collisions
of Metastable Helium Atoms with Rubidium Atoms'',
Tech. Phys. Lett.~{\bf 34}, pp.~693--695 (2008)

\bibitem[ESO web]{ESO_Web}
ESO Laser Systems Department website\\
\href{http://www.eso.org/sci/facilities/develop/lgsf/}
{http://www.eso.org/sci/facilities/develop/lgsf/}

\bibitem[2008]{HWM07}
Drob, D. P., \emph{et al.}, ``An Empirical Model of the Earth's Horizontal
Wind Fields: HWM07'', J.~Geophys.\ Res.~{\bf 113}, doi:10.1029/2008JA01366
(2008)
\href{http://nssdcftp.gsfc.nasa.gov/models/atmospheric/hwm07/}
 {http://nssdcftp.gsfc.nasa.gov/models/atmospheric/hwm07/}

\bibitem[2007]{Drummond}
Drummond, J.; Denman, C.; Hillman, P.; Telle, J.; Eichkoff, M.;
Novotny, S., ``The Sodium LGS Brightness Model over the SOR'', 2007 AMOS
Conference Proceedings of the Advanced Maui Optical and Space Surveillance
Technologies Conference, Wailea, Maui, Hawaii, September 1215, 2007, Ed.: S.
Ryan, The Maui Economic Development Board, p.~E67

\bibitem[2005]{Fishbane}
Fishbane, P. M., Gasiorowicz, S. G., Thornton, S. T., 2005,
Physics for Scientists and Engineers,
Prentice-Hall, Upper Saddle River.

\bibitem[1966]{Fried}
Fried, D. L., ``Optical Resolution Through a Randomly Inhomogeneous Medium for Very Long and Very Short Exposures,'' J.\ Opt.\ Soc.\ Am.\ {\bf 56},
pp.~1372--1379 (1966)
\href{http://www.opticsinfobase.org/abstract.cfm?URI=josa-56-10-1372}{http://www.opticsinfobase.org/abstract.cfm?URI=josa-56-10-1372}

\bibitem[1972]{Happer72}
Happer, W., ``Optical Pumping'', Rev.\ Mod.\ Phys.~{\bf 44}, pp.~169--249
(1972)

\bibitem[1987]{Happer87}
Happer, W. and Van Wijngaarden, W. A.,
``An optical pumping primer'', Hyperfine Interactions {\bf 38},
pp.~435--470 (1987)

\bibitem[1994]{Happer94}
Happer, W., MacDonald, G. J., Max, C. E., and Dyson, F. J.,
``Atmospheric-turbulence compensation by resonant optical backscattering from
the sodium layer in the upper atmosphere'',
J.\ Opt.\ Soc.\ Am.\ A {\bf 11}, pp.~263--276 (1994)

\bibitem[2009]{HapperPriv}
Happer, W., private communication (2009)

\bibitem[2008]{Hillman08}
Hillman, P. D.; Drummond, J. D.; Denman, C. A.; Fugate, R. Q.
``Simple model, including recoil, for the brightness of sodium guide stars
created from CW single frequency fasors and comparison to measurements'',
Adaptive Optics Systems. Edited by Hubin, Norbert; Max, Claire E.;
Wizinowich, Peter L. Proceedings of the SPIE {\bf 7015},
pp.~70150L-70150L-13 (2008)
\href{http://dx.doi.org/10.1117/12.790650}{http://dx.doi.org/10.1117/12.790650}

\bibitem[2008a]{Holz08Marseille}
Holzl\"ohner, R., Bonaccini Calia, D. and Hackenberg, W., ``Physical Optics
Modeling and Optimization of Laser Guide Star Propagation'', Adaptive Optics
Systems, edited by Hubin, N.; Max, Claire E.; Wizinowich, P. L.,
Proceedings of the SPIE {\bf 7015}, pp.~701521-701521-11 (2008)
\href{http://dx.doi.org/10.1117/12.790907}{http://dx.doi.org/10.1117/12.790907}

\bibitem[2008b]{HolzCfAO08}
Holzl\"ohner, R., Bonaccini Calia D. and Hackenberg W., ``Sodium Return Flux
Simulations: Exciter'', Center for Adaptive Optics, talk at
Fall Retreat 2008 (2008)

\bibitem[2009]{Holz09TRE}
Holzl\"ohner, R., Bonaccini Calia D. and Hackenberg W., ``Sodium LGS Return Flux Study'', ESO Technical Report VLT-TRE-ESO-11875-3940, Issue 1 (available upon request) (2009)

\bibitem[2009]{Hubin09}
Hubin, N., ``Review of AO systems studied for the E-ELT'',
Proceedings of AO for ELT conference, abstract 118 (2009)

\bibitem[IGRF 2005]{IGRF}
IGRF Release 2005 geomagnetic model online at NOAA website
\href{http://www.ngdc.noaa.gov/IAGA/vmod/}{http://www.ngdc.noaa.gov/IAGA/vmod/}

\bibitem[1998]{Kartoshkin98}
Kartoshkin, V. A., ``Chemoionization and spin exchange in collisions of excited metastable helium atoms with sodium atoms in the ground state: II.~The calculation of cross
sections for chemoionization and spin exchange'', Opt.\ and Spectr.~{\bf 85},
pp.177--180 (1998)

\bibitem[2009]{Kartoshkin09}
Kartoshkin, V. A., private communication (2009)

\bibitem[2007a]{KibblewhiteRingberg1}
Kibblewhite, E., ``Why understanding the physics of the sodium atom is
important'', Talk at Ringberg Conference, October 2007.
\href{http://www.mpia.de/PARSEC/Ring2007/TalksPostersPDF/Monday/SodiumPhysicsEdKibbleWhite.pdf}
{http://www.mpia.de/PARSEC/Ring2007/TalksPostersPDF/ Monday/SodiumPhysicsEdKibbleWhite.pdf}

\bibitem[2007b]{KibblewhiteRingberg2}
Kibblewhite, E., ``Upgrades to the pulsed sum frequency laser operated on the
5m at Palomar'', Talk at Ringberg Conference, October 2007.
\href{http://www.mpia.de/PARSEC/Ring2007/TalksPostersPDF/Thursday/PalomarLaserUpgrade_EdKibblewhite.pdf}
{http://www.mpia.de/PARSEC/Ring2007/TalksPostersPDF/Thursday/ PalomarLaserUpgrade\_EdKibblewhite.pdf}

\bibitem[2008]{KibRep}
Kibblewhite, E., ``Sodium Laser Guide Star Return
Flux Study for The European Southern Observatory'', contract study for ESO,
E-TRE-KIB-297-0001, 26 August 2008 (available upon request)

\bibitem[1992]{Milonni92}
Milonni, P. W. and Thode, L. E.,
``Theory of mesospheric sodium fluorescence excited by pulse trains'',
Appl.\ Opt.~{\bf 31}, pp.~785--800 (1992)

\bibitem[1998]{Milonni98}
Milonni, P. W., Fugate, R. Q., and Telle, J. M., ``Analysis of measured
photon returns from sodium beacons,'' J.\ Opt.\ Soc.\ Am.\ A {\bf 15},
pp.~217--233 (1998)
\href{http://www.opticsinfobase.org/josaa/abstract.cfm?URI=josaa-15-1-217}{http://www.opticsinfobase.org/josaa/abstract.cfm?URI=josaa-15-1-217}

\bibitem[1999]{Milonni99}
Milonni, P. W., Fearn, H., Telle, J. M. and Fugate, R. Q.,
``Theory of continuous-wave excitation of the sodium beacon'',
J.\ Opt.\ Soc.\ Am.\ A~{\bf 16}, pp.~2555--2566 (1999)
\href{http://www.opticsinfobase.org/josaa/abstract.cfm?uri=josaa-16-10-2555}{http://www.opticsinfobase.org/josaa/abstract.cfm?uri=josaa-16-10-2555}

\bibitem[1994]{Morris}
Morris, J. R., ``Efficient excitation of a mesospheric sodium laser guide star
by intermediate-duration pulses,'' J.\ Opt.\ Soc.\ Am.\ A {\bf 11},
pp.~832--845 (1994)
\href{http://www.opticsinfobase.org/josaa/abstract.cfm?URI=josaa-11-2-832}{http://www.opticsinfobase.org/josaa/abstract.cfm?URI=josaa-11-2-832}

\bibitem[2009a]{Moussaoui09geomag}
Moussaoui, N., Holzl\"ohner, R., Hackenberg, W. and
Bonaccini Calia, D., ``Dependence of sodium laser guide star photon return
on the geomagnetic field'', Astron.\ Astrophys.\ {\bf 501}, pp.~793--799 (2009)
\href{http://www.aanda.org/index.php?option=article&access=standard&Itemid=129&url=/articles/aa/abs/2009/26/aa11411-08/aa11411-08.html}{[link]}

\bibitem[2009b]{Moussaoui09Na}
Moussaoui, N., Clemesha, B. R., Holzl\"ohner, R., Hackenberg, W. and
Bonaccini Calia, D., ``Statistics of the sodium layer parameters at low
geographic latitude and its impact on the adaptive optics sodium laser guide
star characteristics'', in preparation

\bibitem[1990]{MSISE-90}
MSISE-90 atmospheric model online at NASA website
\href{http://ccmc.gsfc.nasa.gov/modelweb/atmos/msise.html}{http://ccmc.gsfc.nasa.gov/modelweb/atmos/msise.html}

\bibitem[1998]{Nagengast}
Nagengast, W., Nass, A., Grosshauser, C., Rith, K., and Schmidt, F.,
`Relaxation of electron polarization for optically pumped rubidium atoms'
J.\ Appl.\ Phys.~{\bf 83}, pp.~5626 (1998)

\bibitem[2000]{dOrgeville}
d'Orgeville, C., Rigaut, F., and Ellerbroek, B. L.,
``LGS AO photon return simulations and laser requirements for the Gemini LGS
AO program'', Proc. SPIE {\bf 4007}, pp.131--141 (2000)
\href{http://spie.org/x648.html?product_id=390400}{http://spie.org/x648.html?product\_id=390400}

\bibitem[2004]{Patat}
Patat, F.,
``Observing During Bright Time: Tips and Tricks'',
The Messenger {\bf 118}, ESO Publication, pp.~11--14 (2004),
\href{http://www.eso.org/sci/publications/messenger/archive/no.118-dec04/messenger-no118.pdf}
{http://www.eso.org/sci/publications/messenger/archive/no.118-dec04/ messenger-no118.pdf}

\bibitem[2006]{Pique06}
Pique J.-P., Moldovan I. C., and Fesquet, V.,
``Concept for polychromatic laser guide stars: one-photon excitation of the
4P$_{3/2}$ level of a sodium atom'',
J.\ Opt.\ Soc.\ Am.\ A {\bf 23}, pp.~2817--2828 (2006)

\bibitem[2009]{Pfrommer}
Pfrommer, T., P. Hickson, and C.-Y. She, ``A large-aperture sodium fluorescence lidar with very high resolution for mesopause dynamics and adaptive optics studies'', Geophys. Res. Lett. {\bf 36}, L15831, doi:10.1029/2009GL038802.

\bibitem[1964]{Ramsey}
Ramsey, A. T. and Anderson, L. W.,
``Spin relaxation in an optically oriented sodium vapor'',
Il Nuovo Cimento {\bf 32}, pp.~1151--1157 (1964)

\bibitem[1969]{Ressler}
Ressler, N. W., Sands, R. H., Stark, T. E.,
`Measurement of Spin-Exchange Cross Sections for Cs133, Rb87, Rb85, K39, and
Na23', Phys.\ Rev. {\bf 184}, pp.~102--118 (1969)

\bibitem[online]{RochesterADM}
Rochester, S. M., Atomic Density Matrix package for Mathematica, version
09.08.07 or later, available at
\href{http://budker.berkeley.edu/ADM/}{http://budker.berkeley.edu/ADM/}

\bibitem[1979]{Simonich}
Simonich, D. M., Clemesha, B. R. and Kirchhoff, V. W., ``The
mesospheric sodium layer at 23$^\circ$S: Nocturnal and seasonal variations'',
J.\ Geophys.\ Res.\ {\bf 84}, pp.1543--1550 (1979)

\bibitem[2009]{Steck}
Steck, D. A., ``Sodium D Line Data'', version 2.1.1, 30 April 2009,
\href{http://steck.us/alkalidata/}{http://steck.us/alkalidata/}

\bibitem[2009]{TaylorCLEO09}
Taylor, L., Friedenauer, A., Protopopov, V., Feng, Y.,
Bonaccini Calia, D., Karpov, V., Hackenberg, W.,
Holzl\"ohner, R., Clements, W., Hager, M., Lison, F., Kaenders, W.,
`20\,Watt 589\,nm via frequency doubling of coherently
beam combined 2-MHz 1178-nm CW signals amplified in Raman
PM Fiber Amplifiers', postdeadline PDA.7, CLEO 2009, Munich, Germany (2009)
\href{http://www.toptica.com/products/itemlayer/215/2009_confirmed_paper_submitted_CLEO.pdf}{[link]}

\bibitem[2006]{Telle06}
Telle, J., Drummond, J., Denman, C., Hillman, P., Moore, G., Novotny, S., and Fugate, R., ``Studies of a Mesospheric sodium Guidestar Pumped by Continuous-Wave Sum-Frequency Mixing of Two Nd:YAG laser Lines in Lithium Triborate,'' Proceedings of the SPIE, {\bf 6215}, pp.~62150K-1 -– 62150K-10, Atmospheric Propagation III (2006)

\bibitem[2008]{Telle08}
Telle, J., Drummond, J., Hillman, P., Denman, C., ``Simulations of Mesospheric Sodium Guidestar Radiance,'' Proceedings of the SPIE {\bf 6878}, pp.~68780G-1 –- 68780G-12, Atmospheric Propagation of Electromagnetic Waves II (2008)

\bibitem[2008]{Thomas}
Thomas, S. J., Gavel, D., Adkins, S., and Kibrick, B.,
``Analysis of on-sky sodium profile data and implications for LGS AO wavefront sensing'', Proc. SPIE 7015, {\bf 70155L} (2008), DOI:10.1117/12.789661

\bibitem[1989]{Ungar}
Ungar, P. J., Weiss, D. S., Riis, E., and Chu, S., ``Optical molasses and
multilevel atoms: theory,'' J.\ Opt.\ Soc.\ Am.\ B {\bf 6},
pp.~2058--2071 (1989)
\href{http://www.opticsinfobase.org/abstract.cfm?URI=josab-6-11-2058}{http://www.opticsinfobase.org/abstract.cfm?URI=josab-6-11-2058}

\bibitem[1992]{Vorst}
van der Vorst, H. A., ``Bi-CGSTAB: A fast and smoothly converging variant of Bi-CG for the solution of nonsymmetric linear systems'', SIAM J.\ Sci.\ Comp.\ {\bf13}, pp.~631--644 (1992)

\bibitem[2004]{Wright}
Wright, M. R., An Introduction to Chemical Kinetics,
Wiley (2004)




















%
%
%
%
%
%
%
%
%
%

\end{thebibliography}
\end{document}